\documentclass[useAMS,usenatbib]{aa}

\def\draftversion{1} % 0 = clean     1 = draft       2 = referee (clean but with \new commands in bold)
\usepackage{amssymb,latexsym,graphicx,natbib,eufrak,times,amsmath,xspace,ifthen, xcolor}
\usepackage[normalem]{ulem}
\usepackage{orcidlink}

\usepackage{hyperref}
\definecolor{linkblue}{rgb}{0,0.4,0.6}
\hypersetup{colorlinks, linkcolor={linkblue}, citecolor={linkblue}, urlcolor={linkblue}}

\ifthenelse{ \draftversion > 0 }{
  \newcommand{\so}[1]{\color[rgb]{0.6,0,0.6}\sout{#1}\color{black}} % Strikethrough 
} {
  
  \newcommand{\so}[1]{}
}

\ifthenelse{\equal{\draftversion}{1}}{
  \newcommand{\sep}[1]{\par\begin{color}[rgb]{0,0.4,0}\begin{center}\hrule\end{center}\end{color}\par} % SEPARATOR: green horizontal line
  \newcommand{\todo}[1]{\begin{color}{red}\ \ifthenelse{\equal{#1}{}} {$\bullet\bullet\bullet$} {$\bullet$\ #1 $\bullet$}\end{color}} % TODO: red
  \newcommand{\idea}[1]{\begin{color}[rgb]{0,0.4,0}\textit{#1}\end{color}} % IDEA: green
  \newcommand{\sk}[1]{\begin{color}[rgb]{0.6,0,0.6}#1\end{color}} % SKIP: purple
  \newcommand{\toc}{\par\begin{color}[rgb]{0.6,0,0.6}\begin{center}\hrule\vspace{0.5mm}\begingroup\small\let\cleardoublepage\relax\let\clearpage\relax\mytoc\endgroup\vspace{0.5mm}\hrule\end{center}\end{color}\par} % TOC

  }{
  \newsavebox{\trashcan}

  \newcommand{\sep}[1]{}
  \newcommand{\todo}[1]{}
  \newcommand{\idea}[1]{}
  \newcommand{\sk}[1]{}
  \newcommand{\toc}{}

  }
 \makeatletter\newcommand\mytoc{\@starttoc{toc}}\makeatother % set TOC style
\long\def\symbolfootnote[#1]#2{\begingroup%
\def\thefootnote{\fnsymbol{footnote}}\footnote[#1]{#2}\endgroup} 

\newcommand{\eqn}[2][]{Equation#1~\ref{eqn:#2}} %% for plural form, use: \eqn[s]{emc2} and (\ref{eqn:emc3})     to get  Equations (1) and (2)
\newcommand{\fig}[2][]{Figure#1~\ref{fig:#2}}

\newcommand{\sect}[2][]{Section#1~\ref{sec:#2}}

\newcommand{\bb}[1]{\ifmmode \mbox{\boldmath $ #1$} \else  \mbox{\boldmath $#1$} \fi}

\newcommand{\mh}{\ensuremath{\textrm{\,--\,}}}    % long hyphen
\newcommand{\dd}{\ensuremath{\,\mathrm{d}}}       % differential d with space (used in integrals)
\newcommand{\U}[1]{\ensuremath{\mathrm{~#1}}}     % units
\newcommand{\e}[1]{\ensuremath{\times 10^{#1}}}   % x 10^x
\newcommand{\yr}{\U{yr}}
\newcommand{\Myr}{\U{Myr}}          \newcommand{\myr}{\Myr}
\newcommand{\Gyr}{\U{Gyr}}          \newcommand{\gyr}{\Gyr}
\newcommand{\pc}{\U{pc}}
\newcommand{\kpc}{\U{kpc}}
          
\newcommand{\Msun}{\U{M}_{\odot}}   \newcommand{\msun}{\Msun}
\newcommand{\Msunyr}{\Msun\yr^{-1}} \newcommand{\msunyr}{\Msunyr}
\newcommand{\Zsun}{\U{Z}_{\odot}}   \newcommand{\zsun}{\Zsun}
\newcommand{\cc}{\U{cm^{-3}}}
\newcommand{\K}{\U{K}}
    
\newcommand{\kms}{\U{km\ s^{-1}}}

\newcommand{\dex}{\U{dex}}

\newcommand{\afe}{\ensuremath{[\alpha/\mathrm{Fe}]}\xspace}       % [alpha/Fe]
\newcommand{\feh}{\ensuremath{[\mathrm{Fe/H}]}\xspace}       % [Fe/H]

\newcommand{\ramses}{{\sc Ramses}\xspace}

\newcommand{\vintergatan}{{\sc Vintergatan}\xspace}

\newcommand{\sxb}{Observatoire Astronomique de Strasbourg, Universit\'e de Strasbourg, CNRS UMR 7550, F-67000 Strasbourg, France\label{sxb}}
\newcommand{\usias}{University of Strasbourg Institute for Advanced Study, 5 all\'ee du G\'en\'eral Rouvillois, F-67083 Strasbourg, France\label{usias}}

\newcommand{\meudon}{Observatoire de Paris, section de Meudon, GEPI, 5 Place Jules Jannsen, 92195 Meudon, France\label{meudon}}

\newcommand{\aip}{Leibniz-Institut f\"ur Astrophysik Potsdam (AIP), An der Sternwarte 16, D-14482, Potsdam, Germany\label{aip}}
\newcommand{\lund}{Lund Observatory, Division of Astrophysics, Department of Physics, Lund University, Box 43, SE-221 00 Lund, Sweden\label{lund}}

\newcommand{\chalmers}{Department of Space, Earth and Environment, Chalmers University of Technology, SE-41296 Gothenburg, Sweden\label{chalmers}}

\begin{document}
\title{Effects of secular growth and mergers on the evolution of metallicity gradients and azimuthal variations in a Milky~Way-like galaxy}
\titlerunning{Evolution of metallicity gradients}

\author{
	% Florent~Renaud\inst{\ref{sxb},\ref{usias}}\orcidlink{0000-0001-5073-2267} \and friends}
	Florent~Renaud\inst{\ref{sxb},\ref{usias}}\orcidlink{0000-0001-5073-2267} \and
	Bridget~Ratcliffe\inst{\ref{aip}}\orcidlink{0000-0003-1124-7378} \and
	Ivan~Minchev\inst{\ref{aip}}\orcidlink{0000-0002-5627-0355} \and
	Misha~Haywood\inst{\ref{meudon}} \and 
	Paola~Di~Matteo\inst{\ref{meudon}} \and
	Oscar~Agertz\inst{\ref{lund}}\orcidlink{0000-0002-4287-1088} \and
	Alessandro~B.~Romeo\inst{\ref{chalmers}}
	}
   
\institute{\sxb\\\email{florent.renaud@astro.unistra.fr} \and \usias \and  \aip \and \meudon \and \lund \and \chalmers}
% \institute{\sxb\\\email{florent.renaud@astro.unistra.fr} \and \usias}
\authorrunning{Renaud et al.}

\date{Received September 12, 2024; accepted October 29, 2024}

%%%%%%%%%%%%%%%%%%%%%%%%%%%%%%%%%%%%%%%%%%%%%%%%%%%%%%%%%%%%%%%%%%%%%%%%%%%%%%%%%

\abstract{We analyze the evolution of the radial profiles and the azimuthal variations of the stellar metallicities from the \vintergatan simulation of a Milky Way-like galaxy. We find that negative gradients exist as soon as the disk settles at high redshift, and are maintained throughout the long term evolution of the galaxy, including during major merger events. The inside-out growth of the disk and an overall outward radial migration tend to flatten these gradients in time. Major merger events only have a moderate and short-lived imprint on the \feh distributions with almost no radial dependence. The reason lies in the timescale for enrichment in Fe being significantly longer than the duration of the starbursts episodes, themselves slower than dynamical mixing during typical interactions. It results that signatures of major mergers become undetectable in \feh only a few Myr after pericenter passages. We note that considering other tracers like the warm interstellar medium, or monitoring the evolution of the metallicity gradient as a single value instead of a radial full profile could lead to different interpretations, and warn against an oversimplification of this complex problem.}
\maketitle

%%%%%%%%%%%%%%%%%%%%%%%%%%%%%%%%%%%%%%%%%%%%%%%%%%%%%%%%%%%%%%%%%%%%%%%%%%%%%%%%%
\section{Introduction}

Chemical abundances are, in principle, one of the most insightful relics in Galactic archeology, as they freeze in time the physical conditions of the galaxies when stars formed \citep{Tinsley1979}. Yet, uncertainties on stellar ages, on the abundances themselves, the incompleteness of surveys, and the diversity of methods, tracers, and results published greatly complicate the task of decoding this information to retrace the formation history of our Milky Way and of galaxies in general.

The method of chemical tagging attempts to retrieve the birth position of stars within their host galaxy \citep{Freeman2002}, arguing that stars formed in the same cluster (that later dissolved) or in the same region of the host galaxy share similar ages and abundances \citep{Bland2010, Ness2022}. This information would then allow to reconstruct the evolution of the galactic disk \citep[but with mixed validity and success, see the discussions in e.g.,][and references therein]{Ting2015, Hogg2016, Smiljanic2018, Garcia2019, Price2020, Casamiquela2021}. Chemical tagging relies on the assumptions of chemical homogeneity of the clusters' parent gas clouds, and on the ability to distinguish different structures with comparable (but not identical) abundances, which requires sufficiently strong time and spatial (radial and/or azimuthal) chemical variations. While the chemical homogeneity of clusters is relatively robust \citep[e.g.,][]{Bovy2016}, the validity domain of the latter assumption remains debated \citep{Wenger2019, Kreckel2020, Ratcliffe2022}. One of the quantifications of these variations is the gradient of the metallicity radial profiles across disk galaxies.

Many galaxies in the Local Universe, including the Milky Way, exhibit negative metallicity gradients \citep[e.g.,][]{Bresolin2007, Moustakas2010, Sanchez2014, Boeche2014, Anders2014, Donor2020}, which has been connected to the inside-out growth of galactic disks \citep[e.g.,][]{Marino2016, Belfiore2017}. However, other galaxies yield flat or even positive slopes \citep[e.g.,][]{Molina2017, Poetrodjojo2018}, notably at high redshift ($z \gtrsim 1$, \citealt{Wuyts2016, Curti2020, Simons2021, Wang2022, Venturi2024}). Interestingly, to our knowledge, no simulation reproduces positive metallicity gradients, regardless of the numerical method and subgrid physics employed \citep{Acharyya2024}. This discrepancy remains an important challenge, and highlights a critical problem in our understanding of galaxy formation.

The evolution of metallicity gradients is also a key element in establishing the history of galaxies \citep[e.g.,][]{Minchev2018, Frankel2018, Lu2022, Ratcliffe2023, Haywood2024}. For instance, \citet{Lu2022} found a steepening of the metallicity gradient in the Milky Way $\approx 8\mh 11 \Gyr$ ago, which they interpreted as a signature of the last major merger \citep[see also][]{Ratcliffe2023}. In this context, the radial migration of stars \citep{Sellwood2002, Roskar2008, Minchev2010, Minchev2013, Monari2016, Carr2022} severely hinders the reconstruction of the metallicity gradient and its evolution, and the associated chemo-kinematical relations \citep{Pilkington2012, Kubryk2013, Minchev2014, Anders2017, Vincenzo2020, Ratcliffe2023, Haywood2024}. Strong assumptions must then be made, like the azimuthal mixing of the interstellar medium (ISM), and the temporal and radial shapes of the metallicity distributions \citep{Minchev2018}. Numerical simulations of galaxy formation are essential to validate these assumptions, and to help the interpretation of the results.

Moreover, the history of galaxies cannot be restricted to an intrinsic evolution in isolation. At least during an early phase of their history, they grow by accreting external matter, either adiabatically along intergalactic filaments or by merging with their neighbors. The external origin of this material implies different chemical compositions, and thus an alteration of the metallicity of the galaxy. On the one hand, the accretion of gas (either at close-to-pristine metallicity from lightly polluted filaments, or at low metallicity from low-mass satellite galaxies which did not self-enrich as much as their massive accretors) is expected to rapidly reduce the overall abundances of the disk, as evoked in the two-infall model \citep{Chiappini1997, Spitoni2019} to explain the observed diversity of the stellar populations in the Milky Way \citep[e.g.,][]{Haywood2013}. On the other hand, interactions and mergers are also suspected to accelerate the radial mixing of the ISM and the stars, leading to flat metallicity gradients \citep{Kewley2010, Rich2012, Munoz2018}. It is therefore essential to understand and characterize the effects of mergers and secular accretion on the chemical contents of galaxies, and the signature(s) they leave on metallicity gradients. Here again, simulations are key to dissect the relative impacts of the various physical processes on the metallicity distributions.

In the MUGS and MaGICC simulations, \citet{Gibson2013} noted that different subgrid recipes for stellar feedback could lead to opposite conclusions: enhanced feedback induces an efficient redistribution of gas over galactic scales, and thus relatively flat metallicity gradients, while a weaker feedback (without pre-supernova contributions in their study) allows for steep gradients at high redshift. At that time, our understanding of stellar feedback and its implementation in galaxy simulations were still in their pioneer era, and rather than treating their study as a validation of one feedback method over the other, \citet{Gibson2013} wisely emphasized that different subgrid models of the same process could lead to a diversity of results and conclusions.

More recently, \citet{Bellardini2022} found in FIRE simulations a steepening of the metallicity gradient of young stars, which form with an initially flat gradient with important azimuthal variations which fade with cosmic time. \citet{Graf2024} complemented this study by noting an age-dependence of the gradients in these simulations. \citet{Buck2023} reported sudden (temporary) steepenings of the gradients in the warm ISM of the NIHAO-UHD simulation suite, which they attribute to the accretion of low-metallicity gas in the galactic outskirts during merger events at high redshift. The differences between the tracers used in these studies (young stars and warm gas) suggests that the underlying physical process behind the change of metallicity gradient may not be the same in both cases.

In their analysis of the FOGGIE simulations, \citet{Acharyya2024} emphasize the limitations of describing the metal distribution in galaxies with the sole value of a global gradient. For instance, they found step-shaped or more complex non-monotonic radial distributions in \feh, which would not be properly described by a single-value slope. A similar behavior has recently been noted in the present-day stellar populations of the Milky Way, of which metallicity profiles yield distinct features near bar resonances \citep{Haywood2024}. These absences of a smooth, monotonous shape of the metallicity profiles hint towards different processes being active in different regions of the galaxy in setting and altering the metal contents. Furthermore, \citet{Acharyya2024} leveraged the high time-frequency of their simulation output to highlight rapid fluctuations of the gradients, concerning $\sim 30$ to $50\%$ of the total time evolution of galaxies. Such rapid variations hinder the interpretation of data if/when galaxies are observed during a short, non-representative state, far from their average behavior.

Studying the full radial metallicity profile --- rather than a single value to represent the gradient --- allows for the possibility to capture potential azimuthal variations and age specific trends, and thus appears as a necessary step towards a holistic understanding of galaxy evolution. In this paper, we propose to follow this route by analyzing the high resolution cosmological zoom-in simulation of a Milky Way-like galaxy \vintergatan. In the following, we use this simulation to illustrate possible variations of the metallicity distributions, and to connect them to their physical causes. Although we build our analysis on a unique formation scenario, we argue that the generalization of our conclusions could apply to a much broader range of histories, even beyond the Milky Way only.

%%%%%%%%%%%%%%%%%%%%%%%%%%%%%%%%%%%%%%%%%%%%%%%
\section{Method}
\label{sec:method}

We use the \vintergatan simulation presented in \citet{Agertz2021} and \citet{Renaud2021, Renaud2021b}, and summarize the numerical technique employed here. \vintergatan is a cosmological zoom-in simulation of a Milky Way-like galaxy run with the \ramses code \citep{Teyssier2002} with adaptive mesh refinement down to $20 \pc$ in the dense interstellar medium. It includes heating from an ultraviolet background, gas self-shielding, atomic and molecular cooling, and star formation at a local efficiency per free-fall time ($\sim 1 \%$, see \citealt{Padoan2012}) above a density and below a temperature thresholds ($100 \cc$, $100 \K$). The stellar component is then sampled with stellar particles of masses $\sim 5000 \msun$ (referred to as stars in the following, for simplicity). Stellar feedback encompasses injection of energy, momentum, mass and heavy elements from winds, radiation pressure, type II and Ia supernovae. 

\vintergatan uses chemical yields for oxygen and iron from \citet{Woosley2007}, and follow their independent evolutions across the simulation. Oxygen is used as a tracer for the $\alpha$ elements. We compute the abundance ratios of the elements $X$ and $Y$ as
\begin{equation}
[Y/X] = \log\left(\frac{f_Y}{f_X}\frac{m_X\, \epsilon_{\odot, X}}{m_Y\, \epsilon_{\odot, Y}}\right),
\end{equation}
where $f$ is the mass fraction of the element, $m$ is its atomic mass, and $\epsilon_{\odot}$ its solar abundance from \citep{Anders1989}. We stress that uncertainties on chemical yields in models and observations necessarily lead to mismatches between the simulated and observed values \citep[e.g.,][]{Blancato2019, Buck2021}. In \vintergatan, we chose to keep the raw values from the simulation rather than post-processing them (e.g., by re-normalization) based on uncertain models or observations. Therefore, we focus our analyses on the relative variation of the abundances and metal profiles, rather than on their absolute values.

The initial conditions are identical to the model ``m12i'' of \citet{Hopkins2014} and \citet{Wetzel2016}, also used in \citet{Bellardini2022}. However, our different numerical treatments and sub-grid recipes lead to significantly different results and conclusions from these works (as discussed later). The zoom-in volume is initialized at $z=100$ with dark matter particles of mass $3.5 \e{4} \msun$, and gas mass resolution of $7.1\e{3} \msun$, with a gas metallicity of $10^{-3} \zsun$ to mimic the enrichment from unresolved population III stars. This leads to the formation of a dark matter halo of radius $R_{200} = 334 \kpc$ and virial mass $M_{200} = 1.3 \e{12} \msun$ at $z=0$. Due to numerical cost, the high resolution hydrodynamical simulation is stopped at $z=0.17$, i.e., several billion years after that galaxy evolution became slow and secular, and thus when this last snapshot is representative of present-day conditions \citep{Renaud2021}. The quality of the match between the simulation and observational data is discussed in \citet{Agertz2021}. The absence of a bar in \vintergatan may affect our conclusions, as strong and long-lived bars influence gas flows, orbital excitations, resonances, and radial migrations \citep{dimatteo2013}. The strong mismatches between the bar fraction in observations and numerical simulations, and its evolution with redshift are a notorious problem in almost all cosmological simulations, which remains to be understood \citep[see][Kraljic et al. in preparation]{Reddish2022}.

All the following analyses are conducted in cylindrical coordinates, with a selection volume defined as the slice of height $\pm 300 \pc$ centered on the plane orthogonal to the angular momentum vector of the stars, and passing by their center of mass, of the most massive progenitor of the most massive galaxy at the end of the simulation. As such, we start the analysis only after the disk settling, which has been estimated at $z\lesssim 5$ by examining the ratio of rotational and dispersion velocities \citep[see][]{Segovia2022}. The growth of the galaxy is dominated by mergers with massive neighbors, until the last major merger starting at $z\approx 1.2$, i.e., at a look-back time of $9 \gyr$ \citep{Renaud2021}. At later times, the galactic disk grows inside-out in a secular way, together with the alignment of the outer, tilting disk (see \citealt{Renaud2021b} for details).

In this paper, we use robust estimators to conduct statistical analyses on sparse datasets, that is the median (instead of the mean), and the robust standard deviation computed as the median absolute deviation (MAD) divided by the scale factor of the normal distribution 0.6745 \citep[see][for details]{Muller2000, Romeo2023}. 

We consider only the instantaneous radius of stars, and not that of their guiding centers. This has implications on the assessments of the effects of radial migration, as discussed later. We compute the radial profiles by measuring the median metallicity in logarithmically-spaced radial bins so as to have larger bins at larger radii, where the density is lower. We call annulus dispersion $\sigma_{\rm annulus}$ the robust standard deviation of the metallicity of all the stars within such a radial bin $\left\{\feh_R\right\}$, that is
\begin{equation}
	\sigma_{\rm annulus} (R) = \frac{{\rm med}\big(\left| \left\{\feh_R\right\} - {\rm med}\left\{\feh_R\right\} \right|\big)}{0.6745},
	\label{eqn:annulus}
\end{equation}
where ${\rm med}$ is the median.

To evaluate the importance of the azimuthal scatter, we split each radial annulus into angular bins of width $5^{\circ}$. Then, we consider the metallicity of stars within this radial and angular bin $\left\{\feh_{R,\theta}\right\}$, and measure its robust standard deviation around the median of the \emph{entire} annulus:
\begin{equation}
	\sigma_{\rm azimuth} (R, \theta) = \frac{{\rm med}\big(\left| \left\{\feh_{R,\theta}\right\} - {\rm med}\left\{\feh_R\right\} \right|\big)}{0.6745}.
\end{equation}
Azimuthal inhomogeneities exist if some of the values of $\sigma_{\rm azimuth} (R, \theta)$ differ from $\sigma_{\rm annulus} (R)$. Therefore, we introduce the dimensionless anisotropy parameter
\begin{equation}
	\chi(R) = {\rm maximum}\left(\frac{\left|\sigma_{\rm azimuth} (R, \theta)-\sigma_{\rm annulus} (R)\right|}{\sigma_{\rm annulus} (R)}\right).
	\label{eqn:anisotropy}
\end{equation}
This parameter is close to zero when the scatter is isotropic, and of the order unity when the scatter in metallicity is caused by strong azimuthal variations. We note that this parameter projects the 2D anisotropy into the radial dimension. Using a bi-dimensional anisotropy parameter (i.e., depending on both $R$ and $\theta$) would provide better estimates of the anisotropic nature of the systems studied, especially for large $\sigma_{\rm annulus}$, but would also be more subject to low-number statistics. For this reason and for the sake of simplicity, we chose to use the definition of \eqn{anisotropy} to estimate the level of anisotropy.

%%%%%%%%%%%%%%%%%%%%%%%%%%%%%%%%%%%%%%%%%%%%%%%
\section{Results}

%%%%%%%%%%%
\subsection{Long term evolution}

%%%%%%%%%%%
\subsubsection{The situation at different epochs}
\label{sec:situation}

\begin{figure*}
\includegraphics{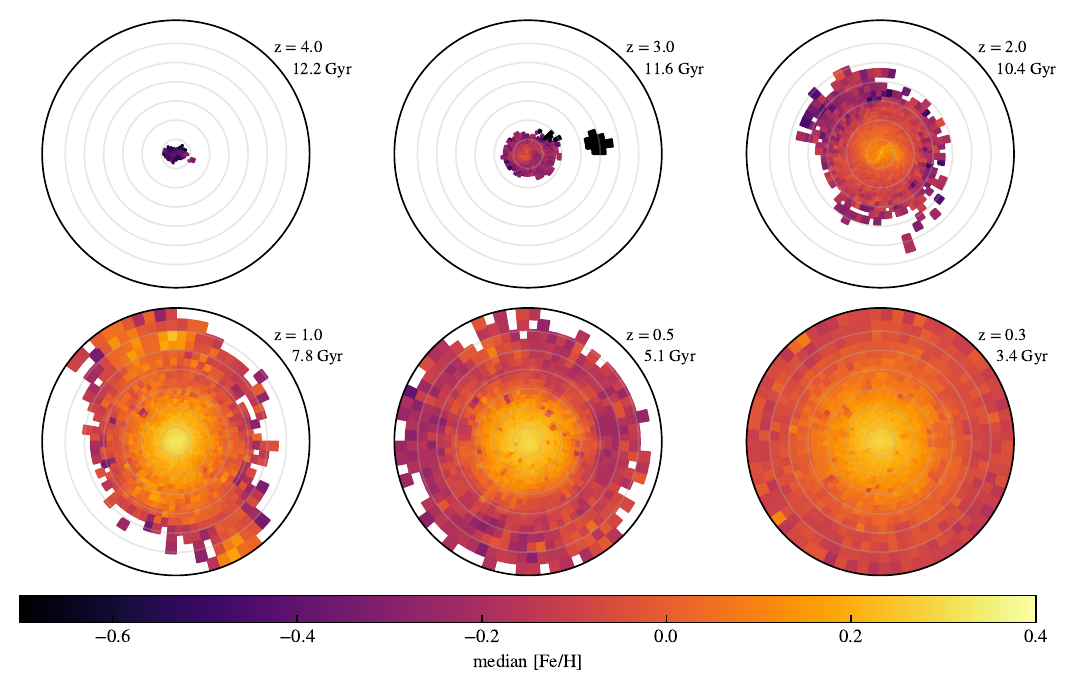}
\caption{Maps of the median metallicity of all the stars in the radial and azimuthal bins described in \sect{method}, at the 6 epochs analyzed in this paper. To ease the comparison between epochs, the color scale is identical for all panels. The radial range covered is $18 \kpc$, with grid lines in gray every $2.5 \kpc$.}
\label{fig:maps}
\end{figure*}

\begin{figure}
\includegraphics{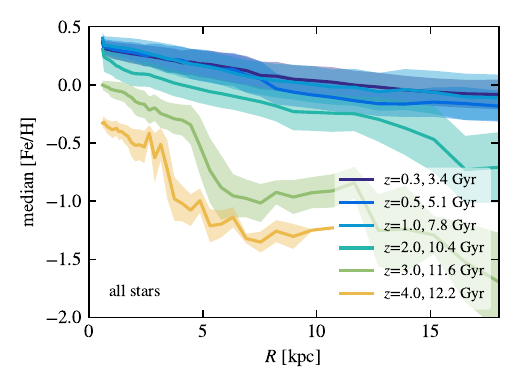}
\caption{Radial profiles of the metallicity of all the stars, measured at different epochs (indicated by their redshifts and look-back times in the legend) as the median in logarithmically-spaced radial bins. The shaded areas show the robust standard deviation $\sigma_{\rm annulus}$ in each radial bin (\eqn{annulus}).}
\label{fig:radprofile}
\end{figure}

\begin{figure}
\includegraphics{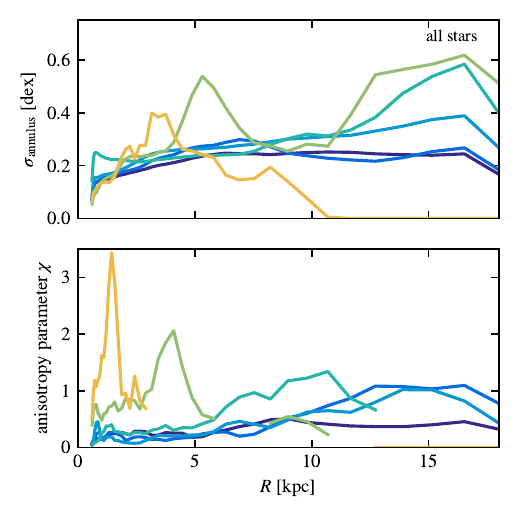}
\caption{Top: robust standard deviation of the metallicity in logarithmically-spaced radial bins. Bottom: anisotropy parameter (\eqn{anisotropy}) quantifying the importance of azimuthal variations in the scatter in metallicity at a given radius. A value $\chi \approx 0$ indicates an isotropic scatter. The colors are as in \fig{radprofile}. Differences in the radial extents of the two quantities are caused by low number statistics in the outer galaxy which forbids the azimuthal binning necessary to compute $\chi$.}
\label{fig:azidisp}
\end{figure}

The median metallicity of all the stars contained within $\pm 300 \pc$ from the galactic mid-plane is mapped in \fig{maps}, at six epochs along the evolution of the galaxy (indicated by redshift and look-back time). A visual inspection already reveals the presence of negative radial gradients at all epochs, including the earliest. However, the azimuthal distributions are complex with fluctuations visible virtually at all radii, and with no clear trend. As stated in \sect{method}, a precise and robust estimate of the 2D anisotropy is a highly non-trivial task, subject to low-number statistics, and we resort instead to a 1D estimate via the $\chi(R)$ parameter for the rest of the paper.

At $z=2$ (top-right panel), the metallicity map reveals the presence of an S-shaped, spiral-like structure composed of metal-rich stars ($\feh \approx 0.0$) in the inner $\approx 2.5 \kpc$. These maps do not contain information on the stellar mass in each bin, and thus this shape does not necessarily correspond to a morphological structure. At this redshift, the galactic disk is too dynamically hot to host the presence of spiral arms (see \citealt{Renaud2021}), likely because of the high gas fraction (\citealt{Renaud2021c}) and repeated tidal stirring by major mergers. Instead, the ISM is organized in a few massive clumps connected by elongated arms, but far from showing the level of the symmetry found here in metallicity. These clumps host the formation of massive clusters in an otherwise relatively smooth and featureless stellar disk. Two massive stellar clumps ($\sim 10^6 \msun$) in particular exist near the extremities of the S-shape. It is possible that stars sharing specific origins (and thus similar kinematics, ages and chemical contents) would respond to the passage of these massive cluster to form the elongated S-shape visible here. The absence of peculiar structures in the maps at other epochs indicates indeed that it is a transient feature created by a precise, and thus rare, configuration.

To quantify the spatial variations in metallicity at these six epochs, \fig{radprofile} shows the radial profiles of the median and robust standard deviation ($\sigma_{\rm annulus}$) of the metallicity of all the stars. ($\sigma_{\rm annulus}$ is also shown in the top panel of \fig{azidisp}.) The earliest epochs in \fig{radprofile} yield a steep gradient of $\approx -0.1 \U{dex\ \kpc^{-1}}$ up to about $3 \mh 4 \kpc$, i.e., approximately 1.5 times the galactic half-mass radius at these times. Further away, the stellar density of the galaxy drops and the few stars beyond this radius are part of small satellites, as visible in the first maps of \fig{maps}. As such, they exhibit a significantly lower \feh with an increased dispersion due to the multiplicity of their origins. At later times, stars found at large radii make the outer galaxy and tidal debris expelled during previous interactions with massive galaxies. The non-linear shape of these distributions, especially at high redshift, reaffirms the cautionary note from \citet{Acharyya2024} that metallicity profiles are not always properly described by a global single-value gradient. A similar conclusion had been reached using \vintergatan by \citet{Agertz2021} who reported radial variations of the present-day metallicity gradients, and also steeper gradients for the low-\afe stars, roughly corresponding to the thin, kinematically cold disk. Here, we add that a clear negative radial gradient in \feh is visible at all epochs and all radii, including before and during the growth phase dominated by mergers ($1 \lesssim z \lesssim 5$): negative gradients are already in place at early times, and are not erased by the repeated galactic interactions. This important result is further discussed in \sect{lmm}. 

The earliest epochs in \fig{radprofile} yield a steep gradient of $\approx -0.1\dex\, \kpc^{-1}$ up to about $3\mh 4 \kpc$, i.e., approximately 1.5 times the galactic half-mass radius ($\approx 2 \kpc$ then). The profiles then flatten with time and it is only after $z=2$ that they reach their final gradient of $\approx -0.05 \dex\, \kpc^{-1}$,\footnote{in good agreement with the observed value of $-0.058 \dex\, \kpc^{-1}$ in the thin disk \citep{Anders2017}.} also up to about 1.5 times the corresponding half-mass radii. This convergence is likely caused by the accelerated star formation around redshift 2, i.e., during the merger-dominated growth phase: only 15\% of the final galactic stellar component is formed before $z=3$, while this fraction jumps to 35\% and 72\% at $z=2$ and 1 respectively, because of repeated merger-induced starbursts. This implies that the statistical weights of features in the metallicity profiles built at the earliest epochs drop when the galaxy forms a lot of stars fast, i.e., during starbursts. This necessarily happens after disk settling ($z < 5$, see \citealt{Segovia2022}), and before the formation of the bulk of the thin, low-\afe disk at $z\approx 1$ (since the merger activity dynamically and chemically prevents the onset of the thin disk, see \citealt{Renaud2021}). Contrarily to the early epochs, breaks and discontinuities are absent from the late radial profiles. Shallower, yet still negative, gradients are found in the outer disk and even beyond ($\approx -0.02\dex\, \kpc^{-1}$ at the latest times).

\fig{azidisp} shows that the dispersion in metallicity in radial annuli $\sigma_{\rm annulus}$ fluctuates with radius at the earliest epochs, but with a moderate amplitude ($\approx 0.2\mh 0.3 \dex$ on average, and $0.6 \dex$ at maximum). However, after $z \approx 1$, it converges towards a rather flat profile. The anisotropy parameter $\chi$ shows similar trends, indicating that the outer part of the galaxies are more prone to azimuthal variations, especially at early times. This results from the influence of the large scale environment of the galaxy, and the slow azimuthal mixing at large radii caused by long dynamical times. Important anisotropies are found over relatively narrow radial ranges at the earliest epochs, translating the irregular morphology of the disk. These trends fade with time, in particular after the last major merger ($z < 1$), where the scatter in metallicity in the disk becomes close to being isotropic ($\chi \ll 1$) over most of the galaxy.

The convergence of the metallicity profiles and gradients (noted after $z= 2$) is more precocious than that of the annulus dispersion ($\sigma_{\rm annulus}$) and anisotropy ($\chi$): the scatters remain high, and with relatively high radial variations until the end of the merger phase ($z=1$). This is particularly pronounced in the outer galaxy, beyond the radial extent of the disk ($\sim 3 \mh 4 \kpc$ at these epochs). The annulus dispersion only really settles in the last two epochs listed here. It monotonically increases with radius in the inner galaxy, to reach a value of $\approx 0.2 \dex$ in the outer galaxy ($\gtrsim 5 \kpc$, i.e., about 90\% of gaseous half-mass radius at the last epochs). These results suggest that, while interactions and mergers do not destroy pre-existing metallicity gradients, they still play an important role in maintaining chemical inhomogeneities across the galactic disks, as explored in \sect{lmm}.

\begin{figure}
\includegraphics{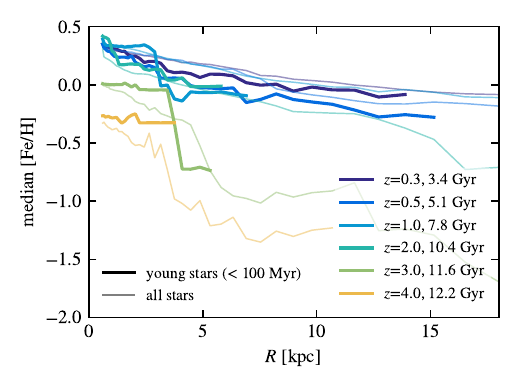}
\caption{Radial profiles of the metallicity of the young stars only. The profiles for all the stars are reproduced from \fig{radprofile} with thin lines, for reference.}
\label{fig:youngradprofile}
\end{figure}

\begin{figure}
\includegraphics{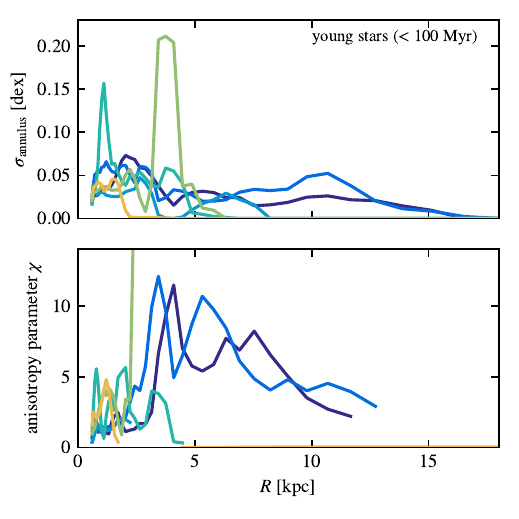}
\caption{Same as \fig{azidisp}, but only for the young stars. Note that the ranges of the vertical axes are not the same as in \fig{azidisp}. The colors are as in \fig[s]{radprofile}, \ref{fig:azidisp}, and \ref{fig:youngradprofile}.}
\label{fig:youngazidisp}
\end{figure}

In \fig[s]{youngradprofile} and \ref{fig:youngazidisp}, we repeat this analysis but limited to stars younger than $100 \Myr$ at each of our six epochs. This naturally strongly limits the impacts of migration of stars from their birth radii (blurring, and churning, including by tidal ejection), and confirms the conclusions presented above. In the radial range where the young and evolved populations overlap, the metallicity of the young stars is, not surprisingly, higher than that of the evolved populations, but only before $z=1$. Later, stars form on average with a lower metallicity than the average of the evolved populations at the same radius. This is due to two distinct effects: (i) in \vintergatan, the highest metallicity is reached during the last major merger at $z = 1.5$: the distribution of \feh reaches a maximum of $\approx 0.7 \dex$. This maximum then slightly decreases to $0.5 \dex$ because of the accretion of low-metallicity gas at the time of the first pericenter of the last major merger\footnote{The exact origin of this gas is difficult to determine, but we suspect its accretion is triggered by the tidal stripping of the incoming satellite galaxy, as other similar patterns of decrease of the maximum metallicity are also seen at the epochs of the first pericenter passages of all major mergers \vintergatan experiences (see \citealt{Renaud2021}, their figures 2 and 4).}. (ii) The absence of mergers at these late epochs facilitates galactic-wide, secular, outward radial migration. The pre-existence of a negative metallicity gradient implies that the migration of metal-rich stars from the inner regions raises the metallicity of the full population at larger radii (compared to the young stars), accompanying the inside-out growth of the disk.

Perhaps not surprisingly, \fig{youngazidisp} shows that the annulus dispersion of the young stars is significantly lower than that of the evolved populations (by factors $\approx 5\mh 10$ on average), as a direct consequence of the more restrictive selection criterion of coeval stars. However, the anisotropy is virtually always higher for the young stars. These variations underline differences in the efficiency of chemical mixing in the cold ISM (traced here by young stars) and in the evolved stellar populations, and different degrees of dynamical coupling between the two components.

The largest anisotropies are found at intermediate radii (i.e., neither in the central-most region, nor in the outer disk), where the ISM is sufficiently dense to form stars, but not perfectly azimuthally mixed. The strongest anisotropies are found at the late epochs at radii $\approx 3\mh 8 \kpc$ because of the stochasticity of the star formation activity ($< 10 \msunyr$ over the entire disk) and possibly also because of the reduced turbulence at this late stage of the disk evolution. The small number of stars concerned (compared to the full disk population) implies that this anisotropy in the young stars does not leave a strong imprint on that of the total population (\fig{azidisp}). 

The opposite situation is found in the inner galaxy ($\lesssim 2 \kpc$), with a high anisotropy early-on which decreases at the late epochs: $\chi$ goes from $2\mh3$ to $1 \mh 1.5$ after $z = 1$. In addition, the annulus dispersion is the highest in this region at all times (except localized, stochastic peaks). This contrasts with the increase of the dispersion with radius noted for all the stars. The short dynamical timescale (crossing time and orbital time) near the galactic center allows for an efficient mixing of the evolved populations. Yet, this effect remains too slow to strongly affect the young stars, which then yield their maximum dispersions in metals in the most active star forming volumes, that is the dense inner galaxy. Particularly at the early epochs, the central region hosts the shortest depletion times of the star-forming gas of the entire galaxy (down to $\sim 100 \Myr$, compared to $\sim 1 \Gyr$ at larger radii). The elevated and complex behavior of the anisotropy parameter there suggests that the large-scale dynamics (kpc scale) are not sufficiently rapid to mix radially and azimuthally the chemical features driven by strongly active and localized star-forming regions\footnote{Although the depletion time, as the ratio of the gas mass and the SFR, reflects the timescale of star formation, it cannot be directly compared to dynamical times, as the true depletion of the gas reservoir is regulated by feedback. Here, we use this estimate to argue that chemical mixing could still be inefficient near the galactic center despite short dynamical timescales, if the star formation timescale is short.}.

At later times, the high dispersion but relatively moderate anisotropy suggest that the orbital mixing (kpc-scale) is more efficient than the turbulent mixing (local) of the ISM. Furthermore, the slowing down of the star formation activity (lower SFR and longer depletion time) now allows for an efficient mixing of star forming gas, which thus tends to homogenize the metallicities of the young stellar populations.

While APOGEE (DR17, \citealt{Abdurro2022}) claims resolutions of $0.02 \dex$ for the disk stars, other spectroscopic surveys like GALAH (DR3, \citealt{Buder2021}) report uncertainties of $\sim 0.1 \dex$. Recently, \citet{Hawkins2023} and \citet{Hackshaw2024} mentioned azimuthal variations in \feh of $\sim 0.1 \dex$ (without age selection). Therefore, the annulus dispersions in \feh of $\approx 0.3 \dex$ (all stars) and $\approx 0.05 \dex$ (young stars only) that we find are close to the current observational limits. This means that at least some of the features of the radial and azimuthal distributions predicted by \vintergatan could not be found before the next generation of surveys.

%%%%%%%%%%%
\subsubsection{Nature versus nurture}
\label{sec:natnut}

\begin{figure}
\includegraphics{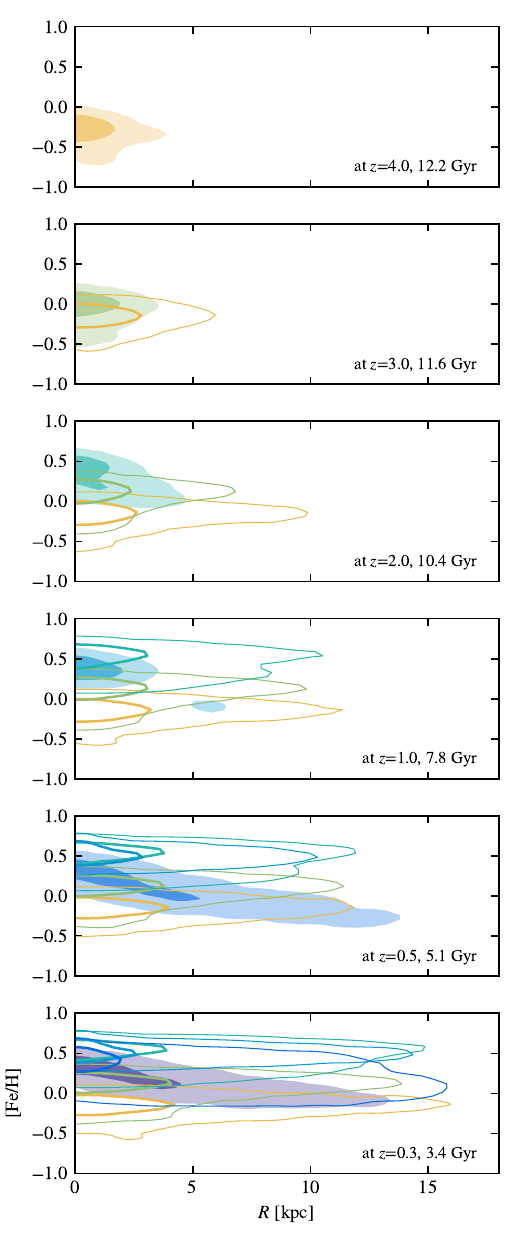}
\caption{Radial distributions of \feh of stars formed at different epochs. Each panel corresponds to one epoch as indicated in the lower-right corner. The filled contours show the distributions of young stars ($< 100 \Myr$) formed at this epoch, with levels corresponding to 0.01 and 1\% of the stellar population. These stars are then tracked to the following epochs, and their evolved distributions are shown in the subsequent panels with line contours, but always keeping the same color across all panels.}
\label{fig:natnut}
\end{figure}

\begin{figure}
\includegraphics[scale=0.96]{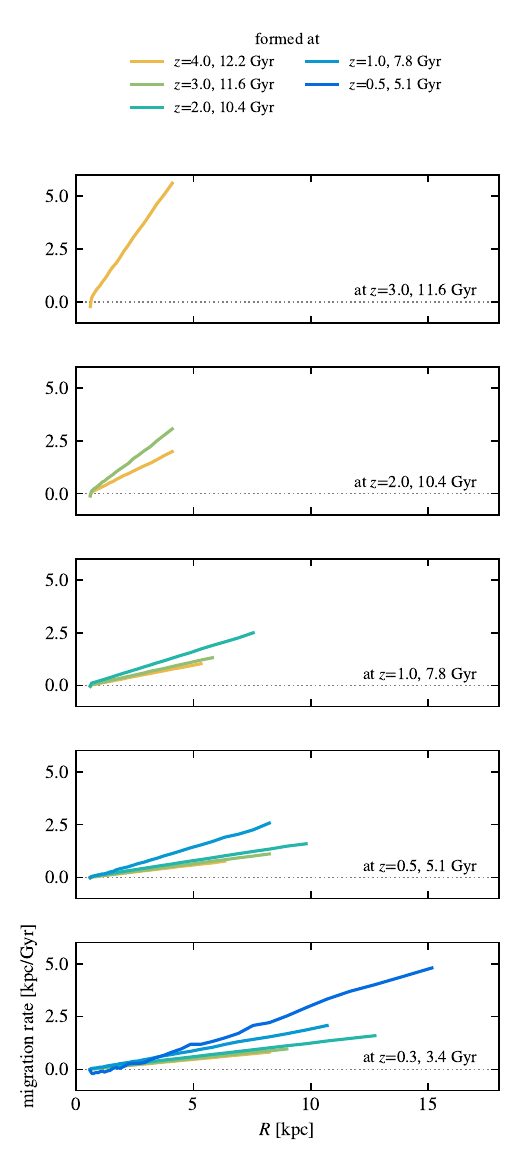}
\caption{Median radial migration rate of stars. As in \fig{natnut}, each color corresponds to a population of stars, followed at different epochs (different panels). The quantity plotted is the median of the difference between the radius at the given epoch (that is the quantity on the $x$-axis) and the birth radius, divided by the time difference between the two epochs, with the convention of a positive value for an outward migration. By considering the instantaneous radius (instead of the average), the migration measured here includes both the blurring and churning effects.}
\label{fig:migration}
\end{figure}

In this section, we establish the relative roles of the conditions at birth and of the dynamical evolution on the metallicity profiles (i.e., nature versus nurture). This is done by selecting young stars formed at different epochs in the galaxy, and tracking them to subsequent times (instead of selecting a new set of young stars from each snapshot as done in the previous section). \fig{natnut} shows the distributions of these (quasi) mono-age populations at birth, selected within $\pm 300 \pc$ from the galactic mid-plane, and with an age $< 100 \Myr$ (filled contours), and then tracked at different stages of the galactic evolution (line contours of the same color in other panels).

Before the last major merger ($z \gtrsim 1$), young stars are confined within a few kpc, but tend to move outward (on average) as they age. However, only a small fraction of the pre-existing stars reach much larger radii, as shown by the absence of significant radial extent of the highest-level contour lines. (The half-mass radius does not significantly vary before the last major merger at $z \approx 1$, see the Figure 6 of \citealt{Agertz2021}.) Therefore, at these early epochs, the radial growth of the thick disk only concerns a few aged stars, and not the young population. Nevertheless, this induces chemical mixing as stars formed in the inner regions can later be found at larger radii. The main inside-out growth of the galaxy occurs after the last major merger, together with the onset of the thin disk (at low \afe), and the alignment of outer titling disk \citep[see][for details]{Renaud2021b}.

The migration rates of the populations are quantified in \fig{migration}. We compute this quantity for every star, as the difference between its galactic radius at a given epoch and its birth radius, divided by the time difference between the two epochs. We then compute the median of this rate in radial bins. In this paper, we use the instantaneous radius, rather than that of the guiding center. As such, our measurements encompasses two effects which modify the galactic radius of stars: blurring (elliptical motions around the guiding center) and churning (increase of the average orbital radius). These effects have different physical causes and respond differently to galactic and environmental effects like bars and mergers exciting orbits. However, distinguishing them is beyond the scope of this paper.

At a given epoch (panel), \fig{migration} shows at which rate a given mono-age population (color) has migrated to its current radius ($x$-axis) since its formation. The median migration is only inward (negative rate) at the last epoch in the innermost $1.5 \kpc$ for the population formed at $z=0.3$ where inward migration reaches $-0.2 \kpc \gyr^{-1}$. Everywhere else and at all the other epochs, the median migration is outwards, with a rate decreasing as the population ages. Furthermore, at a given radius, the migration rate decreases with cosmic time for all the populations. There is no clear evidence for a change of migration regime after the last major merger ($z=1$), suggesting again a moderate role of mergers in the internal dynamics of pre-existing populations, as opposed to the evident and sudden inside-out expansion of star formation shortly after the last major merger $z = 1$.

A direct consequence of the almost ubiquitous outward migration is the reduction of the steepness of the metallicity gradients, also visible in \fig{natnut_gradient}. The metallicity gradients of all the populations at all epochs are either negative or very weak ($\dd \feh / \dd R \lesssim 0$). For the populations of stars formed in the first four epochs ($z \ge 1$), the gradient is negative only in the initial measurement, and null after. This indicates a prompt radial mixing of these populations. However, the stars formed at later times (in the last two panels of \fig{natnut}) still yield a negative gradient after some evolution. These stars form in the thin disk, after the last major merger, and as such experience a less efficient, or at least less rapid, radial mixing. This is also visible in \fig{migration}: the migration rate of the young populations slows down with cosmic time (e.g., the blue curve in the bottom panel is shallower than the orange line in the top panel). The initial gradients tend to be steeper at low redshift, which also contributes to delaying the flattening of the distributions at late times.

%%%%%%%%%%%
\subsection{Evolution during the last major merger}
\label{sec:lmm}

\begin{figure*}
\includegraphics{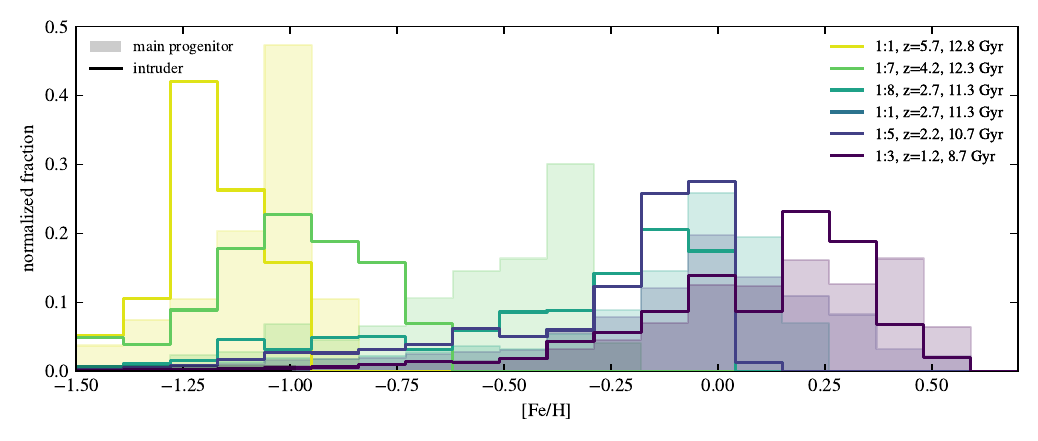}
\caption{Distributions in \feh of all the stars in the galaxies involved in the 6 major merger events in \vintergatan. The distributions are normalized by the total stellar mass of the galaxies. Each color represents a merger event, with the mass ratio, redshift, and look-back time indicated in the top-right corner. Note that, at $z=2.7$, two distinct events cannot be separated by the time resolution of the simulation outputs, and are thus labeled with the same redshift. The \feh distributions of \vintergatan (i.e. the most massive, Milky Way-like progenitor) at the epochs of the mergers are shown with the filled histogram, for comparison with that of its companion galaxies (solid lines).}
\label{fig:progenitors}
\end{figure*}

\begin{figure}
\includegraphics{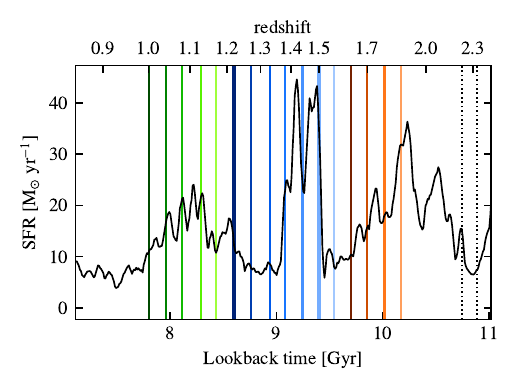}
\includegraphics{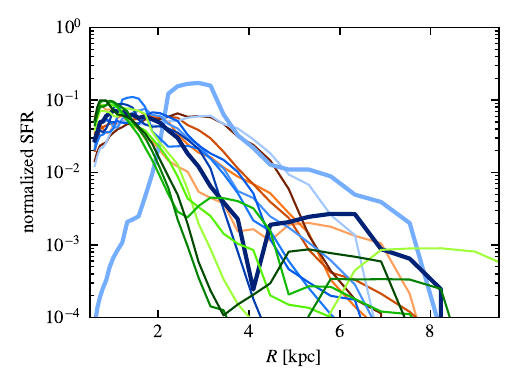}
\caption{Top: in-situ SFR around the epoch of the last major merger, with the colored solid vertical lines indicating the snapshots studied in \sect{lmm}. These snapshots are sorted into three groups: pre-merger (orange), interaction, separation and merger (blue), and post-merger (green). Thick lines indicate the first and second (last) pericenter passages. The dotted lines mark the two pericenter passages of the previous major merger. Bottom: radial distribution of the SFR, normalized to the total SFR at these epochs, and smoothed for the sake of readability.}
\label{fig:lmmsfr}
\end{figure}

\begin{figure}
\includegraphics{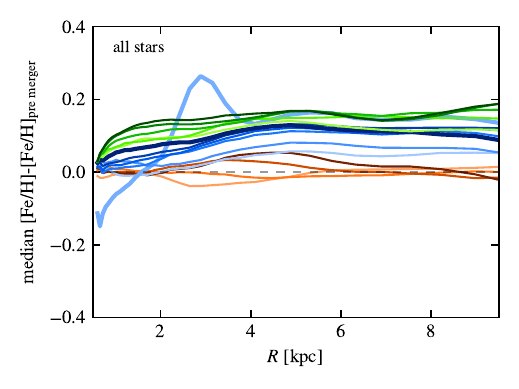}
\caption{Variations of the radial profiles of metallicity of all the star at the instants marked in \fig{lmmsfr}, i.e., approximately every $150 \Myr$ around the epoch of the last major merger. A pre-merger reference profile is computed by stacking the data from the first 3 snapshots listed here, and then subtracted to all the profiles to show the relative variations.}
\label{fig:lmm_evolution}
\end{figure}

\begin{figure}
\includegraphics{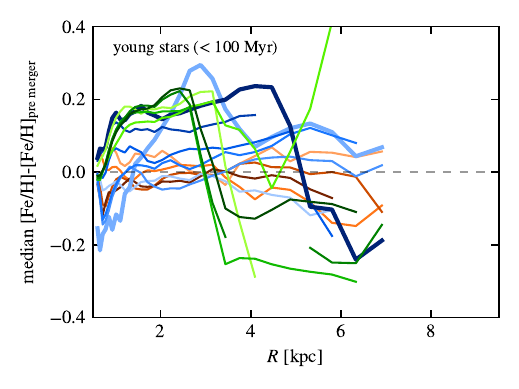}
\caption{Same as \fig{lmm_evolution}, but only for the newly formed stars, i.e., formed less than $100 \Myr$ before each snapshot.}
\label{fig:nslmm_evolution}
\end{figure}

As any other aspect of the galaxy, the distribution of metals is affected by the long-term secular evolution and more punctual and violent events like mergers. For reference, the merger tree of \vintergatan, computed using the stellar component, is shown in \fig{mergertree}. The stochastic nature of mergers and the merger history of the galaxy make it difficult to identify their precise imprint on present-day properties. This is illustrated by the overlap of the metallicity distribution in the progenitor galaxies involved in a major merger (mass ratio > 1:10) shown in \fig{progenitors}. At each merger event, the main galaxy yields a higher maximum metallicity than its lower-mass companion galaxies, as a result of its higher ability to produce and retain metals. For similar reasons, the metal distribution of the main galaxy is systematically wider, reflecting a longer and/or richer formation history. However, by considering galaxies with similar masses (mass ratio > 1:10 in our definition), the differences in the metal contents between the two progenitors of mergers remain within a fraction of dex in all cases. This already suggests that a major merger is unlikely to trigger a drastic change in the metal contents of the main progenitor. Naturally, the opposite situation occurs during minor mergers where the intruder galaxy usually yields significantly poorer metal contents, but the lower mass involved implies a mild contribution to the mass budget of the main, and thus also a limited change of the overall metal contents of the main galaxy.

Keeping these considerations in mind, we study below the effects of major mergers on the pre-existing distributions of metals, and their gradients. To help the interpretation of the simulation data, we focus here on the transformations in the metal distributions caused by the last major merger experienced by \vintergatan. The first passage starts at a look-back time of $8.61 \Gyr$, i.e., $z=1.22$, involving an interloper galaxy of stellar mass of $1.0\e{10} \msun$, i.e., a third of \vintergatan at this epoch, a gas mass of $3.2\e{9} \msun$, and a gas metallicity of $\feh \approx -0.2 \mh 0.6 \dex$. Just before the interaction, both galaxies form stars with $\feh \approx 0\mh 0.5 \dex$, and the half-mass radius of the gaseous component of \vintergatan is $3 \kpc$ \citep{Agertz2021, Renaud2021}. We chose to focus on the last major merger only, as other mergers are closer to each other in time, and their effects could be blended together, making the interpretation more involved.

\fig{lmmsfr} shows the evolution of the in-situ star formation rate (SFR) around the epoch of the last major merger (i.e., the star formation history of the stars formed exclusively in the most massive progenitor of the most massive galaxy at the end of the simulation), and its radial distribution. The first passage triggers a spatially extended burst of star formation, up to $\approx 8 \kpc$, and mainly at radii of $\approx 2\mh 3 \kpc$. During the later stages of the merger and in particular at final coalescence, the boost of SFR is significantly milder, and much more radially concentrated. This concentration is caused by radial gaseous inflows becoming prominent when tidal torques dominate over other triggering mechanisms, i.e., for short separations between the galaxies \citep{Keel1985}. The processes of shocks and tidal and turbulence compressions, which are also known to enhance star formation in mergers and explain off-nuclear starbursts, are generally shorter-lived than the time resolution adopted here \citep{Renaud2014b, Renaud2019}. Therefore, it is possible that their effects are not fully sampled in the radial distributions of \fig{lmmsfr}. The weaker boost of the SFR after the merger than at the first passage could be caused by a partial quenching induced by the interaction (yet, without feedback from an active galactic nucleus, \citealt{Petersson2022}), and notably in the outer galaxy \citep{Moreno2019}, and/or by the long-term depletion of the gas reservoir over timescales of several Gyr (see \citealt{Renaud2021}). The exact reason is out of the scope of the present paper.

Overall, the interaction induces two competing effects on the metal contents: the boost of chemical enrichment caused by the starburst activity, and the mixing of the ISM with metal-poor gas from the interloper itself, and also from the circumgalactic medium of the two galaxies \citep{Olivares2022}. Each effect comes with a time delay with respect to the merger event itself. On the one hand, the intrinsic enrichment is delayed by the timescale of stellar evolution needed to produce and release heavy elements like Fe\footnote{The starburst-induced chemical enrichment is much more rapid when considering lighter elements which are released more promptly by e.g., type-II supernovae. For instance, wet mergers have been shown to leave a clear signature of boosted \afe during the starburst episode they trigger \citep{Renaud2021}.}, followed by the time needed for these ejecta to form the next generation of stars. On the other hand, the diffuse metal-poor gas also takes time to condense (via thermal and dynamical instabilities) and mix with the star forming phase of the ISM (in a similar fashion as the evolution described in \citealt{Semenov2017}). Estimating these delays is a complicated task, and is probably highly dependent on the details of the interaction. We focus here on their combined effects, that is the net chemical evolution throughout the merger.

\fig{lmm_evolution} (respectively \ref{fig:nslmm_evolution}) shows the evolution of the radial metallicity profiles of all the stars (respectively the newly formed stars, younger than $100 \Myr$) before, during, and after the last major merger. The profiles of annulus dispersions and anisotropy parameters are shown in \fig[s]{lmm_evolution_disp} and \ref{fig:nslmm_evolution_disp}. Thanks to the time sampling and volume selection, the stars from the interloper galaxy are only accounted for after the coalescence of the two galaxies, and do not contaminate the measurements in the stages of interaction. For the sake of readability and to enhance the visibility of the variations, a pre-merger reference radial profile computed as a stack of the first three snapshots used here is subtracted from all the profiles. 

\fig{lmm_evolution} shows that the metallicity of the entire stellar population increases gradually, and almost independently of radius during the merger event. In other words, the major merger does not induce, reinforce, or weaken metallicity gradients, as suggested in the previous sections and in contrast to the conclusions of \citet[see \sect{otherworks} for an interpretation of these differences]{Buck2023}. The notable exception is during the first pericenter passage: the metallicity drops in the inner $\approx 1 \kpc$, peaks at $\approx 3 \kpc$, and remains elevated for $\approx 3\mh 10 \kpc$, that is in the outer galaxy. (At this epoch, the half-mass radius of the galaxy is $2\mh3 \kpc$, \citealt{Agertz2021}). Interestingly, this is visible in one snapshot only, indicating that dynamical mixing smooths out these radial variations within less than $2\times 150 \Myr = 300 \Myr$. Later, during the separation phase, we note a mild inflection of the metallicity profile near $3 \kpc$. This approximately corresponds to the co-rotation radius of the interaction, i.e., the radius at which the circular velocity of the main galaxy equals the orbital velocity of the interloper\footnote{This velocity varies along the interaction, between the pericenter and apocenter. We estimate its mean value to be $210 \kms$, giving a co-rotation radius of $\approx 2.5 \kpc$ during the separation phase.}. Due to gravitational torques, galactic matter tends to migrate inwards inside this radius, and outward otherwise, which explains that the metallicity radial profiles yield an inflection caused by different mixing near this radius. This fades after coalescence, once the tidal torques cease. Interestingly, the violent collision at final coalescence does not yield a change as dramatic as for the first passage. No particular radius exhibits strong variations, probably because of efficient mixing across the disk caused by the rapid and direct impact of the interloper after it has lost its orbital angular momentum. 

During the entire interaction episode, the metallicity is globally raised by $0.15 \dex$, that is $\approx 0.08 \dex\, \gyr^{-1}$, except in the innermost region. This rate of enrichment is comparable to that found in the late stages of galaxy evolution ($z \lesssim 2.5$, before $11 \Gyr$ ago, see \citealt{Renaud2021}, their figure 4), but significantly weaker than that in the early phase of rapid galactic growth where it can be as fast as $1.8 \dex\, \gyr^{-1}$. Overall, despite enhancements of the SFR by factors up to 4, the last major merger does not significantly alter the rate of enrichment in \feh across the galaxy. As suggested above, this apparent discrepancy between a starburst activity and an only mild enrichment could be explained by a time delay and/or the accumulation of low-metallicity gas of circumgalactic or intergalactic origin \citep[see e.g.,][]{Renaud2019}.

Because old stars are included in \fig{lmm_evolution}, these profiles gather changes due to the formation during the starburst episodes and the dynamical mixing of the pre-existing stars. To tell apart the two effects, \fig{nslmm_evolution} shows the same measurements, but only considering the stars younger than $100 \Myr$ at the times of the respective snapshots. Here again, the most important enrichment with respect to the pre-merger reference occurs during the interaction phase, near the co-rotation radius ($\approx 2.5 \kpc$). At virtually all stages, the enrichment at large radius ($\gtrsim 4 \kpc$) is slower than in the inner regions, confirming the dilution of the ISM by low-metallicity gas of circumgalactic or intergalactic origin, and the partial dispersal of previously enriched gas outside the co-rotation radius. 

\begin{figure}
\includegraphics{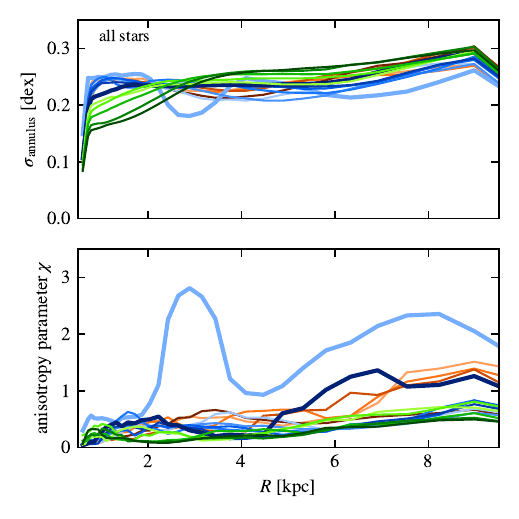}
\caption{Radial profiles of the annulus dispersion and anisotropy parameter for all the stars around the epoch of the last major merger. These quantities correspond to the median profiles in \fig{lmm_evolution}.}
\label{fig:lmm_evolution_disp}
\end{figure}

\begin{figure}
\includegraphics{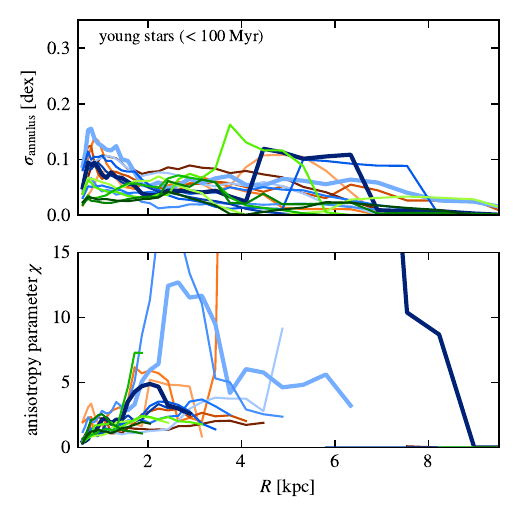}
\caption{Same as \fig{lmm_evolution_disp}, but only for the young stars.}
\label{fig:nslmm_evolution_disp}
\end{figure}

\fig[s]{lmm_evolution_disp} and \ref{fig:nslmm_evolution_disp} show that the merger tends to homogenize the metallicity in the inner galaxy ($\lesssim 2.5 \kpc$), and slightly increase the dispersion in the outer regions. Significant anisotropic features are only visible at the first passage and coalescence (thick lines). When considering all the stars, the anisotropy peaks at the co-rotation ($\approx 2.5\kpc$), likely induced by tidal torques from the interloper, and also well beyond the half-mass radius ($\gtrsim 6\mh 10 \kpc$) corresponding to the ejection of tidal tails made of disk stars in the otherwise metal-poor galactic outskirts. The anisotropy in the metallicity is much stronger for the young stars, especially during the peaks of the SFR (\fig{nslmm_evolution_disp}). However, the timescale of the starburst episodes ($\sim 100\mh 300 \myr$) is significantly shorter than that of enrichment in Fe ($\sim 1 \gyr$ on average, e.g., \citealt{Matteucci2001}), and thus the enhanced star formation cannot leave an immediate imprint on the Fe contents. This conclusion does not necessarily apply to other chemical elements with shorter production and release timescales, like $\alpha$ which can effectively trace starburst episodes (see e.g., \citealt{Renaud2021} and our conclusion below).

Furthermore, these anisotropic features can only be detected in one or two consecutive snapshots, i.e., for less than $\sim 300 \myr$. Since considering all the stars yields comparable timescales, the fading of anisotropy must be caused by efficient dynamical mixing equally affecting the young and old populations. Interestingly, a timescale of $300 \myr$ approximately corresponds to the revolution period of the disk, suggesting that tides from the satellite galaxy could swipe the entire disk over this period, and erase the azimuthal variations in metallicity. This confirms the findings of \citet{Renaud2019} that the short timescales involved during merger-induced starbursts imply a rapid displacement of the star forming volumes across the galactic disks. Therefore, the increased anisotropy at the peaks of SFR is most likely caused by dynamical effects connected to the morphological transformation from tides and the impact of the interloper, rather than enrichment at a specific epoch and position.

In \sect{situation}, we have noted a drop of the azimuthal variations in \feh after the end of the merger phase, and suggested that interactions could maintain chemical inhomogeneities across the disk. The more detailed approach in this section complements this result by adding that this is not caused by a single interaction, nor by the starburst activity. Moderate changes of the dispersion and their rapid decrease rather suggest a limited role of mergers in setting inhomogeneities. We propose that this transition at $z \lesssim 1$ is instead mainly driven by internal, intrinsic factors, possibly the slowing down of the global enrichment rate, and/or the morphological change of the star forming structures set by the lowering of the gas fraction \citep{Renaud2021c}. Other simulations covering a range of formation scenarios are needed to test these hypotheses.

%%%%%%%%%%%%%%%%%%%%%%%%%%%%%%%%%%%%%%%%%%%%%%%
\section{Discussion}

%%%%%%%%%%%%
\subsection{Robustness to variations of the merger history}

Our results are derived from a single galactic formation history, and a single merger event. As such, one can question the range of their validity across all plausible formation scenarios for Milky Way-like galaxies. Yet, \vintergatan experiences six major mergers, naturally with different orbital and intrinsic parameters (mass ratio, orbital inclination, spin-orbit coupling etc.), and we found very little effect on the long term evolution of metallicity distributions over the merger-dominated growth phase. This implies that our conclusions are robust against a change in the merger history of the galaxy. However, in \vintergatan, the last major merger occurs just before the onset of the thin disk: in our scenario, it is indeed the cessation of the merger activity that allows the geometrical, kinematical, and chemical onset of the thin, dynamically cold, and low-\afe disk \citep{Renaud2021}. While this scenario is compatible with spectroscopic data in the Milky Way \citep[see e.g.,][]{Ciuca2020}, one could question what effect a late major merger occurring after the onset of the thin disk would have on the metal distributions. Extrapolating from our conclusions, it is possible that radial migration to large radii would then already be in place in an early thin disk when the last major merger occurs. In these outer regions, interactions are significantly more disruptive, and the timescales are longer than those we have studied above. Therefore, it is possible that an interaction with a pre-existing extended disk would imprint stronger dispersions in the outer galaxy, and that such dispersions would survive for longer than the few $100 \Myr$ we found in the present study. In the Milky Way, the last major merger is thought to have happened $8\mh 10 \gyr$ ago \citep{Nissen2010, Ruchti2015, Haywood2018, Helmi2018, Belokurov2018}, and it is thus very unlikely that the direct signature(s) of its disruption of the metallicity radial profiles could be still detected today.

%%%%%%%%%%%%
\subsection{Absence of a galactic bar}

\vintergatan does not yield a galactic bar, nor an AGN. This probably affects the (re-)distributions of metals in the galactic center caused by resonances and AGN feedback (see e.g., the recent study of \citealt{Verwilghen2024} based on simulations of PHANGS galaxies). Observations reveal that the fraction of strong bars in disk galaxies rapidly decreases with increasing redshift, and seems to level at only $\approx 5\mh 15 \%$ at $z \approx 1 \mh 2$ (see e.g., \citealt{Melvin2014} and references therein, and an update from JWST in \citealt{Leconte2023} at high redshift). If these statistics are representative of the history of the Milky Way, this would suggest that its bar probably did not form before the last major merger \citep[see e.g.,][]{Haywood2024}, and thus only have had a late effect on the metal distributions, confined to the secular growth phase of the Galaxy. By increasing radial migration, a bar would make the metal gradients shallower. In addition, the resonances it creates could induce detectable features like breaks in the metallicity profiles, as predicted by \citet{Khoperskov2020} and found by \citet{Haywood2024}. It is also possible that a bar would dynamically increase the azimuthal dispersions, notably in the older populations experiencing streaming motions from non-axisymmetric structures \citep{dimatteo2013, Grand2016, Orr2023}. Testing these hypotheses requires cosmological simulations that reproduce the bar strength and its correct evolution across cosmic time, which seems to be a challenge for modern simulations (\citealt{Reddish2022}, but see also \citealt{Fragkoudi2020}).

%%%%%%%%%%%%
\subsection{Comparisons with other simulations}
\label{sec:otherworks}

Examining the young stellar populations in the FIRE simulations, \citet{Bellardini2022} reported that the gradients in \feh steepen with time after being initially flat. In \fig[s]{youngradprofile} and \ref{fig:natnut}, we have also noticed that the radial distribution in \feh for the young stars remains almost flat until $z\approx 2$. It is only after the rapid enrichment phase of the galaxy, that the star formation activity gets concentrated in the galactic center, leading to a more rapid increase of the metallicity there than in the outskirts, and thus to a negative gradient in \feh for the young stars. Our results then confirm the findings of \citet{Bellardini2022}. We add that, by including evolved populations, we find that outward motions (migration and tidal effects) naturally stretch the metal distributions with time, which necessarily flattens the gradients for the entire population.

In the NIHAO-UHD simulations, \citet{Buck2023} found that massive mergers induce a steepening of the metallicity gradients. Our analysis however concludes that mergers do not have long-lasting effects on pre-existing gradients. This apparent discrepancy is likely caused by the different approaches we followed. Indeed, \citet{Buck2023} monitor the metallicity of the gas below $1.5\e{4} \K$, while we consider the \feh abundance of stars. Our analysis restricted to the young stellar populations can be connected to the star-forming gas, but important differences remain between the warm ISM (as considered in \citealt{Buck2023}) and the cold, star-forming phase ($< 100 \K$ in our prescription). This is particularly relevant in the low density regions of the galactic outskirts where the cooling times are long \citep{Birnboim2003}, and precisely where \citet{Buck2023} found that mergers deposit the low-metallicity gas which causes the steepening of their gradients. Therefore, before this gas becomes cold and dense enough to form stars, it is possible that it would have mixed with more metal-rich material in the inner disk. The resulting effect on the metallicity gradient of the young stars would then be much less pronounced than that noted by \citet{Buck2023} in the warm ISM, as predicted by our results. In addition, we note that the steepening of the gradients measured in \citet[see their figure 2]{Buck2023} at the times of mergers are short-lived (a few $100 \Myr$), and that the long-term evolution of their galaxies rather tends to yield a flattening of the metallicity profiles, as we also find here.

%%%%%%%%%%%%%%%%%%%%%%%%%%%%%%%%%%%%%%%%%%%%%%%
\section{Summary and conclusion}
\label{sec:conclusion}

Using the \vintergatan cosmological zoom-in simulation of a Milky Way-like galaxy, we have studied the emergence and evolution of stellar metallicity radial distributions in the galactic disk, their gradients, and their dispersion across cosmic time. Our main results are as follows.

\begin{itemize}
\item Negative metallicity gradients are present as soon as the disk settles, and are preserved during the various interactions and mergers the galaxy experiences.
\item Significant azimuthal variations of the metallicities are the most pronounced in the outer regions of the galaxy and at early epochs. They disappear after the last major merger.
\item The median radial migration (blurring and churning) is outwards at all epochs, except in the central $\approx 1.5 \kpc$ of the disk. This migration induces a flattening of the metallicity gradients of mono-age populations with time.
\item The final shape of the metallicity profile of the entire stellar population is reached at $z=2$, even before the last major merger.
\item Young stars exhibit an almost flat metallicity gradient at high redshift. It steepens at later times when the galactic center enriches faster than the outskirts.
\item The migration rate of young stars decreases with time. Moreover, the migration of a given mono-age population also slows down as it ages.
\item Major mergers do not induce, reinforce, or weaken any metallicity gradient. 
\item The metallicity of the entire stellar population gradually increases with time during a major merger, with only weak radial dependences. These are short-lived, moderate variations in \feh in the galactic center (due to tidal-driven inflows) and in the outer galaxy (more sensitive to tidal effects and chemical pollution). The orbital co-rotation radius marks a transition zone.
\item The short-lived nature of these variations in \feh implies that they are not caused by the chemical enrichment from starburst events, but rather by the dynamics of the interaction.
\item Short-lived azimuthal variations in the metallicity profiles are detected after the pericenter passages. Fast azimuthal and radial mixing induced by the interaction erase them in about a rotation period of the disk ($\sim 300 \Myr$).
\item As all the other simulations, we have not reproduced the positive metallicity gradients observed in some galaxies. Explaining the onset of such features remains a challenge for galaxy formation.
\end{itemize}

Our results on the low \feh dispersions and gradients flattening in time validate the assumptions made in methods to determine the birth radius of stars \citep{Minchev2018, Lu2022, Ratcliffe2023}. However, we find that the metallicity gradients of young stars formed at high redshift are almost flat, as a result of enhanced turbulent mixing. This could hinder the assignment of birth radii to the oldest populations ($\gtrsim 11 \Gyr$), although we note that the galactic disk at these epochs spans only a few kpc in radius. We also highlight the peculiar behavior in the galactic center, where intense star formation activity, short dynamical mixing timescales, and inward migration, contrasts with the rest of the galaxy. This further strengthens the cautionary note from \citet{Ratcliffe2024} on the non-representativity of the central metallicity when deriving galactic-wide profiles, and chemo-dynamical evolution scenarios in general.

Our conclusions indicate that the dramatic drop of the dynamical timescales during galactic interactions induces a rapid chemical mixing across the disk(s). Such timescales being shorter than that of enrichment in heavy elements like iron \citep[e.g.,][]{Mennekens2010} and thus even the accelerated release of these elements by the type-Ia supernovae produced during merger-triggered starbursts do not leave a clear, distinct imprint on the distributions of metals. This contrasts with the boost in \afe during starbursts \citep{Renaud2021}, as the lighter $\alpha$ elements are released much faster by type-II supernovae, on timescales comparable or even shorter to the dynamical times of the interaction and/or starburst episode. We then argue that, with sufficient accuracy in the observational chemical and age measurements, future large-scale spectroscopic surveys will be able to identify the epoch of rapid star formation, and associate them with merger events, by searching for their signatures in \afe, but that such an exercise will \emph{always} remain in vain with the use of \feh. Yet, this drawback provides a positive note: the long-term insensitivity to mergers of the iron abundances allows to push back the reconstruction of the assembly of the galactic disk even before the last major merger, and thus to probe up to the earliest epochs of disk settling with appropriate chemical information. The fine details in the radial and azimuthal dispersions of \feh we find are likely to be in reach of the next generation of spectroscopic surveys.

%%%%%%%%%%%%%%%%%%%%%%%%%%%%%%%%%%%%%%%%%%%%%%%%%%%%%%%%%%%%%%%%%%%%%%%%%%%%%%%%
%%%%%%%%%%%%%%%%%%%%%%%%%%%%%%%%%%%%%%%%%%%%%%%%%%%%%%%%%%%%%%%%%%%%%%%%%%%%%%%%
%%%%%%%%%%%%%%%%%%%%%%%%%%%%%%%%%%%%%%%%%%%%%%%%%%%%%%%%%%%%%%%%%%%%%%%%%%%%%%%%
\section*{Acknowledgements}
We thank the referee for their constructive report. FR thanks Nils Hoyer and Katarina Kraljic for helpful discussions, and acknowledges support provided by the University of Strasbourg Institute for Advanced Study (USIAS), within the French national programme Investment for the Future (Excellence Initiative) IdEx-Unistra. BR and IM acknowledge support by the Deutsche Forschungsgemeinschaft under the grant MI 2009/2-1. OA acknowledges support from the Knut and Alice Wallenberg Foundation, the Swedish Research Council (grant 2019-04659), and the Swedish National Space Agency (SNSA Dnr 2023-00164).

\bibliographystyle{aa}
\bibliography{biblio}

\providecommand{\noopsort}[1]{}
\begin{thebibliography}{101}
\expandafter\ifx\csname natexlab\endcsname\relax\def\natexlab#1{#1}\fi

\bibitem[{{Abdurro'uf} {et~al.}(2022){Abdurro'uf}, {Accetta}, {Aerts}, {Silva
  Aguirre}, {Ahumada}, {Ajgaonkar}, {Filiz Ak}, \& {et al.}}]{Abdurro2022}
{Abdurro'uf}, {Accetta}, K., {Aerts}, C., {et~al.} 2022, \apjs, 259, 35

\bibitem[{{Acharyya} {et~al.}(2024){Acharyya}, {Peeples}, {Tumlinson}, {Shea},
  {Lochhaas}, {Wright}, {Simons}, {Augustin}, {Smith}, \& {Hyeonmin
  Lee}}]{Acharyya2024}
{Acharyya}, A., {Peeples}, M.~S., {Tumlinson}, J., {et~al.} 2024, arXiv
  e-prints, arXiv:2404.06613

\bibitem[{{Agertz} {et~al.}(2021){Agertz}, {Renaud}, \& et~al.}]{Agertz2021}
{Agertz}, O., {Renaud}, F., \& et~al. 2021, \mnras, 503, 5826

\bibitem[{{Anders} \& {Grevesse}(1989)}]{Anders1989}
{Anders}, E. \& {Grevesse}, N. 1989, \gca, 53, 197

\bibitem[{{Anders} {et~al.}(2017){Anders}, {Chiappini}, {Rodrigues}, {Miglio},
  {Montalb{\'a}n}, {Mosser}, {Girardi}, {Valentini}, {Noels}, {Morel},
  {Johnson}, {Schultheis}, {Baudin}, {de Assis Peralta}, {Hekker},
  {Theme{\ss}l}, {Kallinger}, {Garc{\'\i}a}, {Mathur}, {Baglin}, {Santiago},
  {Martig}, {Minchev}, {Steinmetz}, {da Costa}, {Maia}, {Allende Prieto},
  {Cunha}, {Beers}, {Epstein}, {Garc{\'\i}a P{\'e}rez},
  {Garc{\'\i}a-Hern{\'a}ndez}, {Harding}, {Holtzman}, {Majewski},
  {M{\'e}sz{\'a}ros}, {Nidever}, {Pan}, {Pinsonneault}, {Schiavon},
  {Schneider}, {Shetrone}, {Stassun}, {Zamora}, \& {Zasowski}}]{Anders2017}
{Anders}, F., {Chiappini}, C., {Rodrigues}, T.~S., {et~al.} 2017, \aap, 597,
  A30

\bibitem[{{Anders} {et~al.}(2014){Anders}, {Chiappini}, {Santiago},
  {Rocha-Pinto}, {Girardi}, {da Costa}, {Maia}, {Steinmetz}, {Minchev},
  {Schultheis}, {Boeche}, {Miglio}, {Montalb{\'a}n}, {Schneider}, {Beers},
  {Cunha}, {Allende Prieto}, {Balbinot}, {Bizyaev}, {Brauer}, {Brinkmann},
  {Frinchaboy}, {Garc{\'\i}a P{\'e}rez}, {Hayden}, {Hearty}, {Holtzman},
  {Johnson}, {Kinemuchi}, {Majewski}, {Malanushenko}, {Malanushenko},
  {Nidever}, {O'Connell}, {Pan}, {Robin}, {Schiavon}, {Shetrone}, {Skrutskie},
  {Smith}, {Stassun}, \& {Zasowski}}]{Anders2014}
{Anders}, F., {Chiappini}, C., {Santiago}, B.~X., {et~al.} 2014, \aap, 564,
  A115

\bibitem[{{Belfiore} {et~al.}(2017){Belfiore}, {Maiolino}, {Tremonti},
  {S{\'a}nchez}, {Bundy}, {Bershady}, {Westfall}, {Lin}, {Drory}, {Boquien},
  {Thomas}, \& {Brinkmann}}]{Belfiore2017}
{Belfiore}, F., {Maiolino}, R., {Tremonti}, C., {et~al.} 2017, \mnras, 469, 151

\bibitem[{{Bellardini} {et~al.}(2022){Bellardini}, {Wetzel}, {Loebman}, \&
  {Bailin}}]{Bellardini2022}
{Bellardini}, M.~A., {Wetzel}, A., {Loebman}, S.~R., \& {Bailin}, J. 2022,
  \mnras, 514, 4270

\bibitem[{{Belokurov} {et~al.}(2018){Belokurov}, {Erkal}, {Evans}, {Koposov},
  \& {Deason}}]{Belokurov2018}
{Belokurov}, V., {Erkal}, D., {Evans}, N.~W., {Koposov}, S.~E., \& {Deason},
  A.~J. 2018, \mnras, 478, 611

\bibitem[{{Birnboim} \& {Dekel}(2003)}]{Birnboim2003}
{Birnboim}, Y. \& {Dekel}, A. 2003, \mnras, 345, 349

\bibitem[{{Blancato} {et~al.}(2019){Blancato}, {Ness}, {Johnston}, {Rybizki},
  \& {Bedell}}]{Blancato2019}
{Blancato}, K., {Ness}, M., {Johnston}, K.~V., {Rybizki}, J., \& {Bedell}, M.
  2019, \apj, 883, 34

\bibitem[{{Bland-Hawthorn} {et~al.}(2010){Bland-Hawthorn}, {Krumholz}, \&
  {Freeman}}]{Bland2010}
{Bland-Hawthorn}, J., {Krumholz}, M.~R., \& {Freeman}, K. 2010, \apj, 713, 166

\bibitem[{{Boeche} {et~al.}(2014){Boeche}, {Siebert}, {Piffl}, {Just},
  {Steinmetz}, {Grebel}, {Sharma}, {Kordopatis}, {Gilmore}, {Chiappini},
  {Freeman}, {Gibson}, {Munari}, {Siviero}, {Bienaym{\'e}}, {Navarro},
  {Parker}, {Reid}, {Seabroke}, {Watson}, {Wyse}, \& {Zwitter}}]{Boeche2014}
{Boeche}, C., {Siebert}, A., {Piffl}, T., {et~al.} 2014, \aap, 568, A71

\bibitem[{{Bovy} {et~al.}(2016){Bovy}, {Rix}, {Schlafly}, {Nidever},
  {Holtzman}, {Shetrone}, \& {Beers}}]{Bovy2016}
{Bovy}, J., {Rix}, H.-W., {Schlafly}, E.~F., {et~al.} 2016, \apj, 823, 30

\bibitem[{{Bresolin}(2007)}]{Bresolin2007}
{Bresolin}, F. 2007, \apj, 656, 186

\bibitem[{{Buck} {et~al.}(2023){Buck}, {Obreja}, {Ratcliffe}, {Lu}, {Minchev},
  \& {Macci{\`o}}}]{Buck2023}
{Buck}, T., {Obreja}, A., {Ratcliffe}, B., {et~al.} 2023, \mnras, 523, 1565

\bibitem[{{Buck} {et~al.}(2021){Buck}, {Rybizki}, {Buder}, {Obreja},
  {Macci{\`o}}, {Pfrommer}, {Steinmetz}, \& {Ness}}]{Buck2021}
{Buck}, T., {Rybizki}, J., {Buder}, S., {et~al.} 2021, \mnras, 508, 3365

\bibitem[{{Buder} {et~al.}(2021){Buder}, {Sharma}, {Kos}, {Amarsi},
  {Nordlander}, {Lind}, {Martell}, {Asplund}, \& {et al.}}]{Buder2021}
{Buder}, S., {Sharma}, S., {Kos}, J., {et~al.} 2021, \mnras, 506, 150

\bibitem[{{Carr} {et~al.}(2022){Carr}, {Johnston}, {Laporte}, \&
  {Ness}}]{Carr2022}
{Carr}, C., {Johnston}, K.~V., {Laporte}, C. F.~P., \& {Ness}, M.~K. 2022,
  \mnras, 516, 5067

\bibitem[{{Casamiquela} {et~al.}(2021){Casamiquela}, {Castro-Ginard}, {Anders},
  \& {Soubiran}}]{Casamiquela2021}
{Casamiquela}, L., {Castro-Ginard}, A., {Anders}, F., \& {Soubiran}, C. 2021,
  \aap, 654, A151

\bibitem[{{Chiappini} {et~al.}(1997){Chiappini}, {Matteucci}, \&
  {Gratton}}]{Chiappini1997}
{Chiappini}, C., {Matteucci}, F., \& {Gratton}, R. 1997, \apj, 477, 765

\bibitem[{{Ciuc{\u{a}}} {et~al.}(2020){Ciuc{\u{a}}}, {Kawata}, {Miglio},
  {Davies}, \& {Grand}}]{Ciuca2020}
{Ciuc{\u{a}}}, I., {Kawata}, D., {Miglio}, A., {Davies}, G.~R., \& {Grand}, R.
  J.~J. 2020, arXiv e-prints, arXiv:2003.03316

\bibitem[{{Curti} {et~al.}(2020){Curti}, {Mannucci}, {Cresci}, \&
  {Maiolino}}]{Curti2020}
{Curti}, M., {Mannucci}, F., {Cresci}, G., \& {Maiolino}, R. 2020, \mnras, 491,
  944

\bibitem[{{Di Matteo} {et~al.}(2013){Di Matteo}, {Haywood}, {Combes},
  {Semelin}, \& {Snaith}}]{dimatteo2013}
{Di Matteo}, P., {Haywood}, M., {Combes}, F., {Semelin}, B., \& {Snaith}, O.~N.
  2013, \aap, 553, A102

\bibitem[{{Donor} {et~al.}(2020){Donor}, {Frinchaboy}, {Cunha}, {O'Connell},
  {Allende Prieto}, {Almeida}, {Anders}, {Beaton}, {Bizyaev}, {Brownstein},
  {Carrera}, {Chiappini}, {Cohen}, {Garc{\'\i}a-Hern{\'a}ndez}, {Geisler},
  {Hasselquist}, {J{\"o}nsson}, {Lane}, {Majewski}, {Minniti}, {Bidin}, {Pan},
  {Roman-Lopes}, {Sobeck}, \& {Zasowski}}]{Donor2020}
{Donor}, J., {Frinchaboy}, P.~M., {Cunha}, K., {et~al.} 2020, \aj, 159, 199

\bibitem[{{Fragkoudi} {et~al.}(2020){Fragkoudi}, {Grand}, {Pakmor},
  {Bl{\'a}zquez-Calero}, {Gargiulo}, {Gomez}, {Marinacci}, {Monachesi}, {Ness},
  {Perez}, {Tissera}, \& {White}}]{Fragkoudi2020}
{Fragkoudi}, F., {Grand}, R.~J.~J., {Pakmor}, R., {et~al.} 2020, \mnras, 494,
  5936

\bibitem[{{Frankel} {et~al.}(2018){Frankel}, {Rix}, {Ting}, {Ness}, \&
  {Hogg}}]{Frankel2018}
{Frankel}, N., {Rix}, H.-W., {Ting}, Y.-S., {Ness}, M., \& {Hogg}, D.~W. 2018,
  \apj, 865, 96

\bibitem[{{Freeman} \& {Bland-Hawthorn}(2002)}]{Freeman2002}
{Freeman}, K. \& {Bland-Hawthorn}, J. 2002, \araa, 40, 487

\bibitem[{{Garcia-Dias} {et~al.}(2019){Garcia-Dias}, {Allende Prieto},
  {S{\'a}nchez Almeida}, \& {Alonso Palicio}}]{Garcia2019}
{Garcia-Dias}, R., {Allende Prieto}, C., {S{\'a}nchez Almeida}, J., \& {Alonso
  Palicio}, P. 2019, \aap, 629, A34

\bibitem[{{Gibson} {et~al.}(2013){Gibson}, {Pilkington}, {Brook}, {Stinson}, \&
  {Bailin}}]{Gibson2013}
{Gibson}, B.~K., {Pilkington}, K., {Brook}, C.~B., {Stinson}, G.~S., \&
  {Bailin}, J. 2013, \aap, 554, A47

\bibitem[{{Graf} {et~al.}(2024){Graf}, {Wetzel}, {Bellardini}, \&
  {Bailin}}]{Graf2024}
{Graf}, R.~L., {Wetzel}, A., {Bellardini}, M.~A., \& {Bailin}, J. 2024, arXiv
  e-prints, arXiv:2402.15614

\bibitem[{{Grand} {et~al.}(2016){Grand}, {Springel}, {Kawata}, {Minchev},
  {S{\'a}nchez-Bl{\'a}zquez}, {G{\'o}mez}, {Marinacci}, {Pakmor}, \&
  {Campbell}}]{Grand2016}
{Grand}, R. J.~J., {Springel}, V., {Kawata}, D., {et~al.} 2016, \mnras, 460,
  L94

\bibitem[{{Hackshaw} {et~al.}(2024){Hackshaw}, {Hawkins}, {Filion}, {Horta},
  {Laporte}, {Carr}, \& {Price-Whelan}}]{Hackshaw2024}
{Hackshaw}, Z., {Hawkins}, K., {Filion}, C., {et~al.} 2024, arXiv e-prints,
  arXiv:2405.18120

\bibitem[{{Hawkins}(2023)}]{Hawkins2023}
{Hawkins}, K. 2023, \mnras, 525, 3318

\bibitem[{{Haywood} {et~al.}(2013){Haywood}, {Di Matteo}, {Lehnert}, {Katz}, \&
  {G{\'o}mez}}]{Haywood2013}
{Haywood}, M., {Di Matteo}, P., {Lehnert}, M.~D., {Katz}, D., \& {G{\'o}mez},
  A. 2013, \aap, 560, A109

\bibitem[{{Haywood} {et~al.}(2018){Haywood}, {Di Matteo}, {Lehnert}, {Snaith},
  {Khoperskov}, \& {G{\'o}mez}}]{Haywood2018}
{Haywood}, M., {Di Matteo}, P., {Lehnert}, M.~D., {et~al.} 2018, \apj, 863, 113

\bibitem[{{Haywood} {et~al.}(2024){Haywood}, {Khoperskov}, {Cerqui}, {Di
  Matteo}, {Katz}, \& {Snaith}}]{Haywood2024}
{Haywood}, M., {Khoperskov}, S., {Cerqui}, V., {et~al.} 2024, arXiv e-prints,
  arXiv:2403.08963

\bibitem[{{Helmi} {et~al.}(2018){Helmi}, {Babusiaux}, {Koppelman}, {Massari},
  {Veljanoski}, \& {Brown}}]{Helmi2018}
{Helmi}, A., {Babusiaux}, C., {Koppelman}, H.~H., {et~al.} 2018, \nat, 563, 85

\bibitem[{{Hogg} {et~al.}(2016){Hogg}, {Casey}, {Ness}, {Rix},
  {Foreman-Mackey}, {Hasselquist}, {Ho}, {Holtzman}, {Majewski}, {Martell},
  {M{\'e}sz{\'a}ros}, {Nidever}, \& {Shetrone}}]{Hogg2016}
{Hogg}, D.~W., {Casey}, A.~R., {Ness}, M., {et~al.} 2016, \apj, 833, 262

\bibitem[{{Hopkins} {et~al.}(2014){Hopkins}, {Kere{\v s}}, {O{\~n}orbe},
  {Faucher-Gigu{\`e}re}, {Quataert}, {Murray}, \& {Bullock}}]{Hopkins2014}
{Hopkins}, P.~F., {Kere{\v s}}, D., {O{\~n}orbe}, J., {et~al.} 2014, \mnras,
  445, 581

\bibitem[{{Keel} {et~al.}(1985){Keel}, {Kennicutt}, {Hummel}, \& {van der
  Hulst}}]{Keel1985}
{Keel}, W.~C., {Kennicutt}, Jr., R.~C., {Hummel}, E., \& {van der Hulst}, J.~M.
  1985, \aj, 90, 708

\bibitem[{{Kewley} {et~al.}(2010){Kewley}, {Rupke}, {Zahid}, {Geller}, \&
  {Barton}}]{Kewley2010}
{Kewley}, L.~J., {Rupke}, D., {Zahid}, H.~J., {Geller}, M.~J., \& {Barton},
  E.~J. 2010, \apjl, 721, L48

\bibitem[{{Khoperskov} {et~al.}(2020){Khoperskov}, {Di Matteo}, {Haywood},
  {G{\'o}mez}, \& {Snaith}}]{Khoperskov2020}
{Khoperskov}, S., {Di Matteo}, P., {Haywood}, M., {G{\'o}mez}, A., \& {Snaith},
  O.~N. 2020, \aap, 638, A144

\bibitem[{{Kreckel} {et~al.}(2020){Kreckel}, {Ho}, {Blanc}, {Glover}, {Groves},
  {Rosolowsky}, {Bigiel}, {Boqu{\'\i}en}, {Chevance}, {Dale}, {Deger},
  {Emsellem}, {Grasha}, {Kim}, {Klessen}, {Kruijssen}, {Lee}, {Leroy}, {Liu},
  {McElroy}, {Meidt}, {Pessa}, {Sanchez-Blazquez}, {Sandstrom}, {Santoro},
  {Scheuermann}, {Schinnerer}, {Schruba}, {Utomo}, {Watkins}, \&
  {Williams}}]{Kreckel2020}
{Kreckel}, K., {Ho}, I.~T., {Blanc}, G.~A., {et~al.} 2020, \mnras, 499, 193

\bibitem[{{Kubryk} {et~al.}(2013){Kubryk}, {Prantzos}, \&
  {Athanassoula}}]{Kubryk2013}
{Kubryk}, M., {Prantzos}, N., \& {Athanassoula}, E. 2013, \mnras, 436, 1479

\bibitem[{{Le Conte} {et~al.}(2023){Le Conte}, {Gadotti}, {Ferreira},
  {Conselice}, {de S{\'a}-Freitas}, {Kim}, {Neumann}, {Fragkoudi},
  {Athanassoula}, \& {Adams}}]{Leconte2023}
{Le Conte}, Z.~A., {Gadotti}, D.~A., {Ferreira}, L., {et~al.} 2023, arXiv
  e-prints, arXiv:2309.10038

\bibitem[{{Lu} {et~al.}(2022){Lu}, {Minchev}, {Buck}, {Khoperskov},
  {Steinmetz}, {Libeskind}, {Cescutti}, \& {Freeman}}]{Lu2022}
{Lu}, Y., {Minchev}, I., {Buck}, T., {et~al.} 2022, arXiv e-prints,
  arXiv:2212.04515

\bibitem[{{Marino} {et~al.}(2016){Marino}, {Gil de Paz}, {S{\'a}nchez},
  {S{\'a}nchez-Bl{\'a}zquez}, {Cardiel}, {Castillo-Morales}, {Pascual},
  {V{\'\i}lchez}, {Kehrig}, {Moll{\'a}}, {Mendez-Abreu},
  {Catal{\'a}n-Torrecilla}, {Florido}, {Perez}, {Ruiz-Lara}, {Ellis},
  {L{\'o}pez-S{\'a}nchez}, {Gonz{\'a}lez Delgado}, {de Lorenzo-C{\'a}ceres},
  {Garc{\'\i}a-Benito}, {Galbany}, {Zibetti}, {Cortijo}, {Kalinova}, {Mast},
  {Iglesias-P{\'a}ramo}, {Papaderos}, {Walcher}, \&
  {Bland-Hawthorn}}]{Marino2016}
{Marino}, R.~A., {Gil de Paz}, A., {S{\'a}nchez}, S.~F., {et~al.} 2016, \aap,
  585, A47

\bibitem[{{Matteucci} \& {Recchi}(2001)}]{Matteucci2001}
{Matteucci}, F. \& {Recchi}, S. 2001, \apj, 558, 351

\bibitem[{{Melvin} {et~al.}(2014){Melvin}, {Masters}, {Lintott}, {Nichol},
  {Simmons}, {Bamford}, {Casteels}, {Cheung}, {Edmondson}, {Fortson},
  {Schawinski}, {Skibba}, {Smith}, \& {Willett}}]{Melvin2014}
{Melvin}, T., {Masters}, K., {Lintott}, C., {et~al.} 2014, \mnras, 438, 2882

\bibitem[{{Mennekens} {et~al.}(2010){Mennekens}, {Vanbeveren}, {De Greve}, \&
  {De Donder}}]{Mennekens2010}
{Mennekens}, N., {Vanbeveren}, D., {De Greve}, J.~P., \& {De Donder}, E. 2010,
  \aap, 515, A89

\bibitem[{{Minchev} {et~al.}(2018){Minchev}, {Anders}, {Recio-Blanco},
  {Chiappini}, {de Laverny}, {Queiroz}, {Steinmetz}, {Adibekyan}, {Carrillo},
  {Cescutti}, {Guiglion}, {Hayden}, {de Jong}, {Kordopatis}, {Majewski},
  {Martig}, \& {Santiago}}]{Minchev2018}
{Minchev}, I., {Anders}, F., {Recio-Blanco}, A., {et~al.} 2018, \mnras, 481,
  1645

\bibitem[{{Minchev} {et~al.}(2013){Minchev}, {Chiappini}, \&
  {Martig}}]{Minchev2013}
{Minchev}, I., {Chiappini}, C., \& {Martig}, M. 2013, \aap, 558, A9

\bibitem[{{Minchev} {et~al.}(2014){Minchev}, {Chiappini}, {Martig},
  {Steinmetz}, {de Jong}, {Boeche}, {Scannapieco}, {Zwitter}, {Wyse}, {Binney},
  {Bland -Hawthorn}, {Bienaym{\'e}}, {Famaey}, {Freeman}, {Gibson}, {Grebel},
  {Gilmore}, {Helmi}, {Kordopatis}, {Lee}, {Munari}, {Navarro}, {Parker},
  {Quillen}, {Reid}, {Siebert}, {Siviero}, {Seabroke}, {Watson}, \&
  {Williams}}]{Minchev2014}
{Minchev}, I., {Chiappini}, C., {Martig}, M., {et~al.} 2014, \apjl, 781, L20

\bibitem[{{Minchev} \& {Famaey}(2010)}]{Minchev2010}
{Minchev}, I. \& {Famaey}, B. 2010, \apj, 722, 112

\bibitem[{{Molina} {et~al.}(2017){Molina}, {Ibar}, {Swinbank}, {Sobral},
  {Best}, {Smail}, {Escala}, \& {Cirasuolo}}]{Molina2017}
{Molina}, J., {Ibar}, E., {Swinbank}, A.~M., {et~al.} 2017, \mnras, 466, 892

\bibitem[{{Monari} {et~al.}(2016){Monari}, {Famaey}, {Siebert}, {Grand},
  {Kawata}, \& {Boily}}]{Monari2016}
{Monari}, G., {Famaey}, B., {Siebert}, A., {et~al.} 2016, \mnras, 461, 3835

\bibitem[{{Moreno} {et~al.}(2019){Moreno}, {Torrey}, {Ellison}, {Patton},
  {Hopkins}, {Bueno}, {Hayward}, {Narayanan}, {Kere{\v{s}}}, {Bluck}, \&
  {Hernquist}}]{Moreno2019}
{Moreno}, J., {Torrey}, P., {Ellison}, S.~L., {et~al.} 2019, \mnras, 485, 1320

\bibitem[{{Moustakas} {et~al.}(2010){Moustakas}, {Kennicutt}, {Tremonti},
  {Dale}, {Smith}, \& {Calzetti}}]{Moustakas2010}
{Moustakas}, J., {Kennicutt}, Robert~C., J., {Tremonti}, C.~A., {et~al.} 2010,
  \apjs, 190, 233

\bibitem[{{Mu{\~n}oz-Elgueta} {et~al.}(2018){Mu{\~n}oz-Elgueta},
  {Torres-Flores}, {Amram}, {Hernandez-Jimenez}, {Urrutia-Viscarra}, {Mendes de
  Oliveira}, \& {G{\'o}mez-L{\'o}pez}}]{Munoz2018}
{Mu{\~n}oz-Elgueta}, N., {Torres-Flores}, S., {Amram}, P., {et~al.} 2018,
  \mnras, 480, 3257

\bibitem[{M{\"u}ller(2000)}]{Muller2000}
M{\"u}ller, J. 2000, Journal of Research of the National Institute of Standards
  and Technology, 105, 551

\bibitem[{{Ness} {et~al.}(2022){Ness}, {Wheeler}, {McKinnon}, {Horta}, {Casey},
  {Cunningham}, \& {Price-Whelan}}]{Ness2022}
{Ness}, M.~K., {Wheeler}, A.~J., {McKinnon}, K., {et~al.} 2022, \apj, 926, 144

\bibitem[{{Nissen} \& {Schuster}(2010)}]{Nissen2010}
{Nissen}, P.~E. \& {Schuster}, W.~J. 2010, \aap, 511, L10

\bibitem[{{Olivares} {et~al.}(2022){Olivares}, {Salom{\'e}}, {Hamer}, {Combes},
  {Gaspari}, {Kolokythas}, {O'Sullivan}, {Beckmann}, {Babul}, {Polles},
  {Lehnert}, {Loubser}, {Donahue}, {Gendron-Marsolais}, {Lagos}, {Pineau des
  Forets}, {Godard}, {Rose}, {Tremblay}, {Ferland}, \&
  {Guillard}}]{Olivares2022}
{Olivares}, V., {Salom{\'e}}, P., {Hamer}, S.~L., {et~al.} 2022, \aap, 666, A94

\bibitem[{{Orr} {et~al.}(2023){Orr}, {Burkhart}, {Wetzel}, {Hopkins}, {Escala},
  {Strom}, {Goldsmith}, {Pineda}, {Hayward}, \& {Loebman}}]{Orr2023}
{Orr}, M.~E., {Burkhart}, B., {Wetzel}, A., {et~al.} 2023, \mnras, 521, 3708

\bibitem[{{Padoan} {et~al.}(2012){Padoan}, {Haugb{\o}lle}, \&
  {Nordlund}}]{Padoan2012}
{Padoan}, P., {Haugb{\o}lle}, T., \& {Nordlund}, {\r{A}}. 2012, \apjl, 759, L27

\bibitem[{{Petersson} {et~al.}(2023){Petersson}, {Renaud}, {Agertz}, {Dekel},
  \& {Duc}}]{Petersson2022}
{Petersson}, J., {Renaud}, F., {Agertz}, O., {Dekel}, A., \& {Duc}, P.-A. 2023,
  \mnras, 518, 3261

\bibitem[{{Pilkington} {et~al.}(2012){Pilkington}, {Few}, {Gibson}, {Calura},
  {Michel-Dansac}, {Thacker}, {Moll{\'a}}, {Matteucci}, {Rahimi}, {Kawata},
  {Kobayashi}, {Brook}, {Stinson}, {Couchman}, {Bailin}, \&
  {Wadsley}}]{Pilkington2012}
{Pilkington}, K., {Few}, C.~G., {Gibson}, B.~K., {et~al.} 2012, \aap, 540, A56

\bibitem[{{Poetrodjojo} {et~al.}(2018){Poetrodjojo}, {Groves}, {Kewley},
  {Medling}, {Sweet}, {van de Sande}, {Sanchez}, {Bland-Hawthorn}, {Brough},
  {Bryant}, {Cortese}, {Croom}, {L{\'o}pez-S{\'a}nchez}, {Richards}, {Zafar},
  {Lawrence}, {Lorente}, {Owers}, \& {Scott}}]{Poetrodjojo2018}
{Poetrodjojo}, H., {Groves}, B., {Kewley}, L.~J., {et~al.} 2018, \mnras, 479,
  5235

\bibitem[{{Price-Jones} {et~al.}(2020){Price-Jones}, {Bovy}, {Webb}, {Allende
  Prieto}, {Beaton}, {Brownstein}, {Cohen}, {Cunha}, {Donor}, {Frinchaboy},
  {Garc{\'\i}a-Hern{\'a}ndez}, {Lane}, {Majewski}, {Nidever}, \&
  {Roman-Lopes}}]{Price2020}
{Price-Jones}, N., {Bovy}, J., {Webb}, J.~J., {et~al.} 2020, \mnras, 496, 5101

\bibitem[{{Ratcliffe} {et~al.}(2024){Ratcliffe}, {Khoperskov}, {Minchev}, {Lu},
  {de Jong}, \& {Steinmetz}}]{Ratcliffe2024}
{Ratcliffe}, B., {Khoperskov}, S., {Minchev}, I., {et~al.} 2024, arXiv
  e-prints, arXiv:2401.09260

\bibitem[{{Ratcliffe} {et~al.}(2023){Ratcliffe}, {Minchev}, {Anders},
  {Khoperskov}, {Guiglion}, {Buck}, {Cunha}, {Queiroz}, {Nitschelm},
  {Meszaros}, {Steinmetz}, {de Jong}, {Nepal}, {Lane}, \&
  {Sobeck}}]{Ratcliffe2023}
{Ratcliffe}, B., {Minchev}, I., {Anders}, F., {et~al.} 2023, \mnras, 525, 2208

\bibitem[{{Ratcliffe} {et~al.}(2022){Ratcliffe}, {Ness}, {Buck}, {Johnston},
  {Sen}, {Beraldo e Silva}, \& {Debattista}}]{Ratcliffe2022}
{Ratcliffe}, B.~L., {Ness}, M.~K., {Buck}, T., {et~al.} 2022, \apj, 924, 60

\bibitem[{{Reddish} {et~al.}(2022){Reddish}, {Kraljic}, {Petersen}, {Tep},
  {Dubois}, {Pichon}, {Peirani}, {Bournaud}, {Choi}, {Devriendt}, {Jackson},
  {Martin}, {Park}, {Volonteri}, \& {Yi}}]{Reddish2022}
{Reddish}, J., {Kraljic}, K., {Petersen}, M.~S., {et~al.} 2022, \mnras, 512,
  160

\bibitem[{{Renaud} {et~al.}(2021{\natexlab{a}}){Renaud}, {Agertz}, {Andersson},
  {Read}, {Ryde}, {Bensby}, {Rey}, \& {Feuillet}}]{Renaud2021b}
{Renaud}, F., {Agertz}, O., {Andersson}, E.~P., {et~al.} 2021{\natexlab{a}},
  \mnras, 503, 5868

\bibitem[{{Renaud} {et~al.}(2021{\natexlab{b}}){Renaud}, {Agertz}, {Read},
  {Ryde}, {Andersson}, {Bensby}, {Rey}, \& {Feuillet}}]{Renaud2021}
{Renaud}, F., {Agertz}, O., {Read}, J.~I., {et~al.} 2021{\natexlab{b}}, \mnras,
  503, 5846

\bibitem[{{Renaud} {et~al.}(2019){Renaud}, {Bournaud}, {Daddi}, \&
  {Wei{\ss}}}]{Renaud2019}
{Renaud}, F., {Bournaud}, F., {Daddi}, E., \& {Wei{\ss}}, A. 2019, \aap, 621,
  A104

\bibitem[{{Renaud} {et~al.}(2014){Renaud}, {Bournaud}, {Kraljic}, \&
  {Duc}}]{Renaud2014b}
{Renaud}, F., {Bournaud}, F., {Kraljic}, K., \& {Duc}, P.-A. 2014, \mnras, 442,
  L33

\bibitem[{{Renaud} {et~al.}(2021{\natexlab{c}}){Renaud}, {Romeo}, \&
  {Agertz}}]{Renaud2021c}
{Renaud}, F., {Romeo}, A.~B., \& {Agertz}, O. 2021{\natexlab{c}}, \mnras, 508,
  352

\bibitem[{{Rich} {et~al.}(2012){Rich}, {Torrey}, {Kewley}, {Dopita}, \&
  {Rupke}}]{Rich2012}
{Rich}, J.~A., {Torrey}, P., {Kewley}, L.~J., {Dopita}, M.~A., \& {Rupke},
  D.~S.~N. 2012, \apj, 753, 5

\bibitem[{{Romeo} {et~al.}(2023){Romeo}, {Agertz}, \& {Renaud}}]{Romeo2023}
{Romeo}, A.~B., {Agertz}, O., \& {Renaud}, F. 2023, \mnras, 518, 1002

\bibitem[{{Ro{\v{s}}kar} {et~al.}(2008){Ro{\v{s}}kar}, {Debattista}, {Quinn},
  {Stinson}, \& {Wadsley}}]{Roskar2008}
{Ro{\v{s}}kar}, R., {Debattista}, V.~P., {Quinn}, T.~R., {Stinson}, G.~S., \&
  {Wadsley}, J. 2008, \apjl, 684, L79

\bibitem[{{Ruchti} {et~al.}(2015){Ruchti}, {Read}, {Feltzing}, {Serenelli},
  {McMillan}, {Lind}, {Bensby}, {Bergemann}, \& {et al.}}]{Ruchti2015}
{Ruchti}, G.~R., {Read}, J.~I., {Feltzing}, S., {et~al.} 2015, \mnras, 450,
  2874

\bibitem[{{S{\'a}nchez} {et~al.}(2014){S{\'a}nchez}, {Rosales-Ortega},
  {Iglesias-P{\'a}ramo}, {Moll{\'a}}, {Barrera-Ballesteros}, {Marino},
  {P{\'e}rez}, {S{\'a}nchez-Blazquez}, \& {et al.}}]{Sanchez2014}
{S{\'a}nchez}, S.~F., {Rosales-Ortega}, F.~F., {Iglesias-P{\'a}ramo}, J.,
  {et~al.} 2014, \aap, 563, A49

\bibitem[{{Segovia Otero} {et~al.}(2022){Segovia Otero}, {Renaud}, \&
  {Agertz}}]{Segovia2022}
{Segovia Otero}, {\'A}., {Renaud}, F., \& {Agertz}, O. 2022, \mnras, 516, 2272

\bibitem[{{Sellwood} \& {Binney}(2002)}]{Sellwood2002}
{Sellwood}, J.~A. \& {Binney}, J.~J. 2002, \mnras, 336, 785

\bibitem[{{Semenov} {et~al.}(2017){Semenov}, {Kravtsov}, \&
  {Gnedin}}]{Semenov2017}
{Semenov}, V.~A., {Kravtsov}, A.~V., \& {Gnedin}, N.~Y. 2017, \apj, 845, 133

\bibitem[{{Simons} {et~al.}(2021){Simons}, {Papovich}, {Momcheva}, {Trump},
  {Brammer}, {Estrada-Carpenter}, {Backhaus}, {Cleri}, {Finkelstein},
  {Giavalisco}, {Ji}, {Jung}, {Matharu}, \& {Weiner}}]{Simons2021}
{Simons}, R.~C., {Papovich}, C., {Momcheva}, I., {et~al.} 2021, \apj, 923, 203

\bibitem[{{Smiljanic} {et~al.}(2018){Smiljanic}, {Donati}, {Bragaglia},
  {Lemasle}, \& {Romano}}]{Smiljanic2018}
{Smiljanic}, R., {Donati}, P., {Bragaglia}, A., {Lemasle}, B., \& {Romano}, D.
  2018, \aap, 616, A112

\bibitem[{{Spitoni} {et~al.}(2019){Spitoni}, {Silva Aguirre}, {Matteucci},
  {Calura}, \& {Grisoni}}]{Spitoni2019}
{Spitoni}, E., {Silva Aguirre}, V., {Matteucci}, F., {Calura}, F., \&
  {Grisoni}, V. 2019, \aap, 623, A60

\bibitem[{{Teyssier}(2002)}]{Teyssier2002}
{Teyssier}, R. 2002, \aap, 385, 337

\bibitem[{{Ting} {et~al.}(2015){Ting}, {Conroy}, \& {Goodman}}]{Ting2015}
{Ting}, Y.-S., {Conroy}, C., \& {Goodman}, A. 2015, \apj, 807, 104

\bibitem[{{Tinsley}(1979)}]{Tinsley1979}
{Tinsley}, B.~M. 1979, \apj, 229, 1046

\bibitem[{{Venturi} {et~al.}(2024){Venturi}, {Carniani}, {Parlanti},
  {Kohandel}, {Curti}, {Pallottini}, {Vallini}, {Arribas}, {Bunker}, {Cameron},
  {Castellano}, {Ferrara}, {Fontana}, {Gallerani}, {Gelli}, {Maiolino},
  {Ntormousi}, {Pacifici}, {Pentericci}, {Salvadori}, \&
  {Vanzella}}]{Venturi2024}
{Venturi}, G., {Carniani}, S., {Parlanti}, E., {et~al.} 2024, arXiv e-prints,
  arXiv:2403.03977

\bibitem[{{Verwilghen} {et~al.}(2024){Verwilghen}, {Emsellem}, {Renaud},
  {Valentini}, {Sun}, {Jeffreson}, {Klessen}, {Sormani}, {Barnes}, {Dolag},
  {Grasha}, {Liang}, {Meidt}, {Neumann}, {Querejeta}, {Schinnerer}, \&
  {Williams}}]{Verwilghen2024}
{Verwilghen}, P., {Emsellem}, E., {Renaud}, F., {et~al.} 2024, \aap, 687, A53

\bibitem[{{Vincenzo} \& {Kobayashi}(2020)}]{Vincenzo2020}
{Vincenzo}, F. \& {Kobayashi}, C. 2020, \mnras, 496, 80

\bibitem[{{Wang} {et~al.}(2022){Wang}, {Jones}, {Vulcani}, {Treu}, {Morishita},
  {Roberts-Borsani}, {Malkan}, {Henry}, {Brammer}, {Strait}, {Brada{\v{c}}},
  {Boyett}, {Calabr{\`o}}, {Castellano}, {Fontana}, {Glazebrook}, {Kelly},
  {Leethochawalit}, {Marchesini}, {Santini}, {Trenti}, \& {Yang}}]{Wang2022}
{Wang}, X., {Jones}, T., {Vulcani}, B., {et~al.} 2022, \apjl, 938, L16

\bibitem[{{Wenger} {et~al.}(2019){Wenger}, {Balser}, {Anderson}, \&
  {Bania}}]{Wenger2019}
{Wenger}, T.~V., {Balser}, D.~S., {Anderson}, L.~D., \& {Bania}, T.~M. 2019,
  \apj, 887, 114

\bibitem[{{Wetzel} {et~al.}(2016){Wetzel}, {Hopkins}, {Kim}, {Faucher-Giguere},
  {Keres}, \& {Quataert}}]{Wetzel2016}
{Wetzel}, A.~R., {Hopkins}, P.~F., {Kim}, J.-h., {et~al.} 2016, ArXiv e-prints
  [\eprint[arXiv]{1602.05957}]

\bibitem[{{Woosley} \& {Heger}(2007)}]{Woosley2007}
{Woosley}, S.~E. \& {Heger}, A. 2007, \physrep, 442, 269

\bibitem[{{Wuyts} {et~al.}(2016){Wuyts}, {Wisnioski}, {Fossati}, {F{\"o}rster
  Schreiber}, {Genzel}, {Davies}, {Mendel}, {Naab}, {R{\"o}ttgers}, {Wilman},
  {Wuyts}, {Bandara}, {Beifiori}, {Belli}, {Bender}, {Brammer}, {Burkert},
  {Chan}, {Galametz}, {Kulkarni}, {Lang}, {Lutz}, {Momcheva}, {Nelson},
  {Rosario}, {Saglia}, {Seitz}, {Tacconi}, {Tadaki}, {{\"U}bler}, \& {van
  Dokkum}}]{Wuyts2016}
{Wuyts}, E., {Wisnioski}, E., {Fossati}, M., {et~al.} 2016, \apj, 827, 74

\end{thebibliography}

%%%%%%%%%%%%%%%%%%%%%%%%%%%%%%%%%%%%%%%%%%%%%%%%%%%%%%%%%%%%%%%%%%%%%%%%%%%%%%%%%%%%%%%%%%%%%%%%%%%%%%%%%%%%%%%%%%%%%%%%%%%%%%%%%%%%%%%%%%%%%
\appendix
\section{Metallicity gradients of mono-age populations}
\label{sec:gradients}

\begin{figure}
\includegraphics{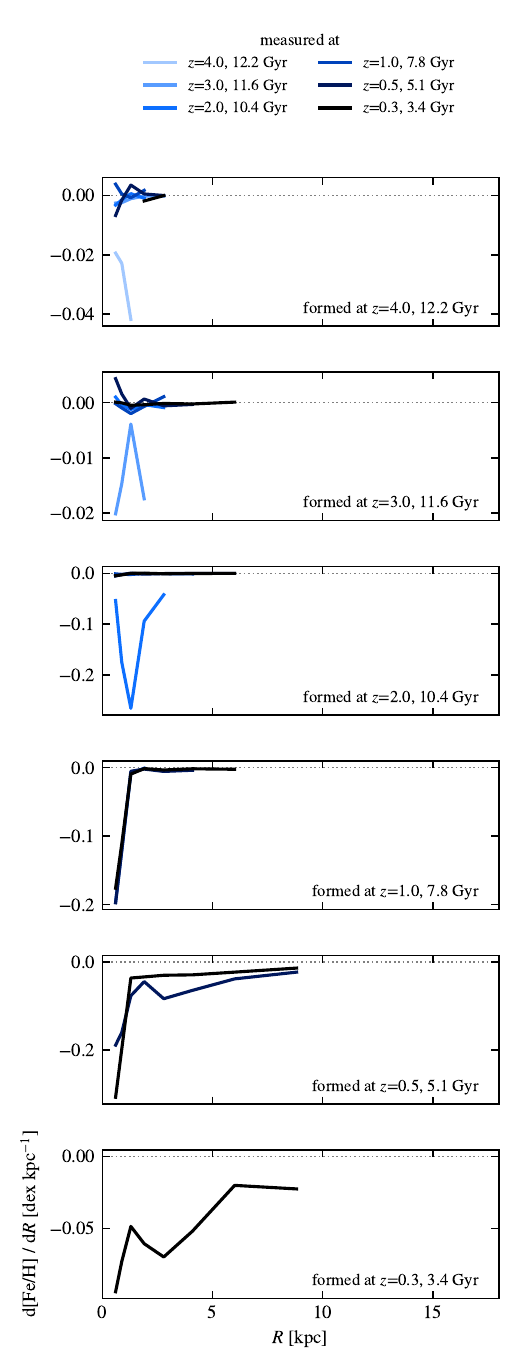}
\caption{Evolution of the radial metallicity gradients for the populations of stars formed at the 6 epochs discussed in \sect{natnut}. Here, each panel corresponds to one population, and the evolution of its gradient is shown with darkening colors.}
\label{fig:natnut_gradient}
\end{figure}

To complement \fig{natnut}, \fig{natnut_gradient} shows the evolution of the metallicity gradients of mono-ages populations (one population per panel), formed at each of the 6 epochs considered in \sect{natnut}. For all formation epochs (i.e., in each panel), the evolution of the population corresponds to a flattening of the gradients in \feh.

\section{Merger tree}
\label{sec:mergertree}

\begin{figure*}
\includegraphics{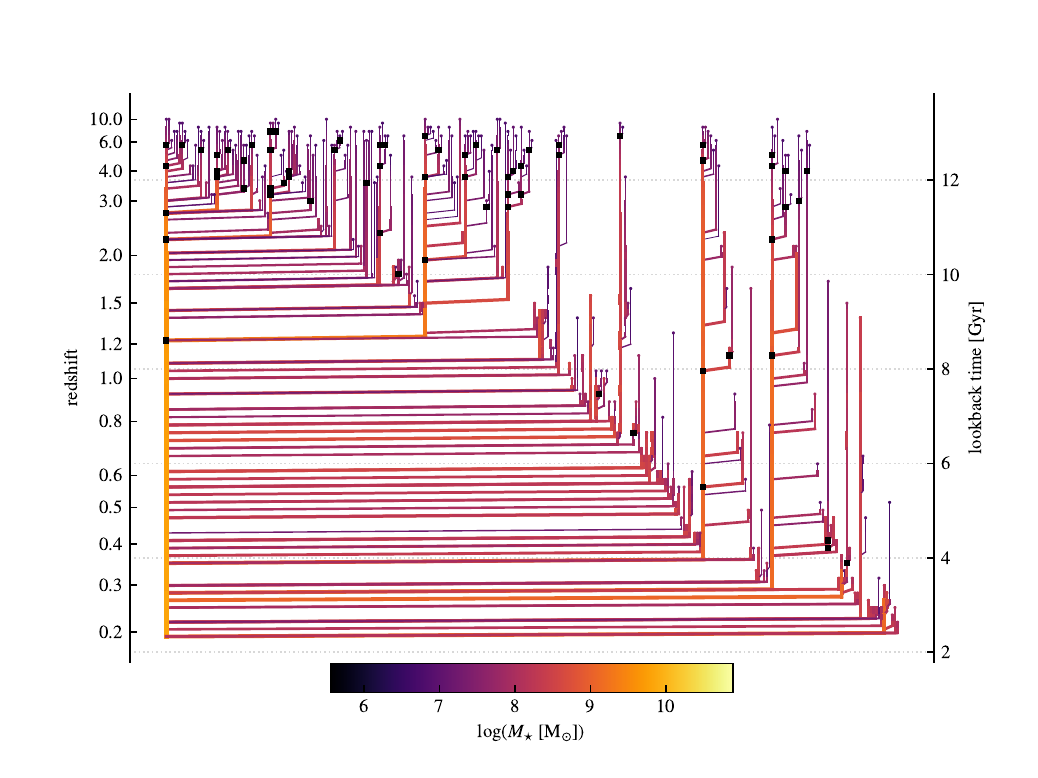}
\caption{Merger tree of \vintergatan, computed using the stellar component. A dot at the top of a branch marks the earliest detection of the stellar component of the corresponding galaxy. The thickness and the color of the branches indicate the stellar mass. Major mergers i.e., mergers with a stellar mass ratio greater than 1:10 are indicated by a black square.}
\label{fig:mergertree}
\end{figure*}

The hierarchical assembly of \vintergatan is illustrated by its merger tree in \fig{mergertree}. The main galaxy is the left-most branch. After an intense bombardment by relatively massive companions until $z\approx 1$, the late build-up of the galaxy is slower, paced by minor mergers only. This transition is key in the construction of the thick versus thin disk, the kinematics, and the chemical contents of these structures, as described in \citet{Renaud2021}.

\end{document}